# Error-Free Demodulation of Pixelated Interferograms: Updated review and circumstantial evidence of plagiarism by 4D Technology Corporation

**MANUEL SERVIN\***

*\*Centro de Investigaciones en Optica A. C., Loma del Bosque 115, Col. Lomas del Campestre, 37150 Leon Guanajuato, Mexico.*
*mservin@cio.mx*
Leon, Mexico, May 2025.

**Abstract:** This paper examines the historical development that led to error-free phase demodulation of pixelated spatial-carrier interferograms between 2004 and 2010. It also evaluates evidence suggesting that 4D Technology Corporation (4DTC), in their SPIE 7790 publication [18], may have adopted key concepts from a manuscript submitted by Servin and collaborators, which at that time was still under peer review [17]. We analyze 4DTC's algorithms based on 2×2 and 3×3 spatial convolution phase-shifting algorithms (SC-PSAs), developed from 2004 to 2009 [9–15], and show that they were fundamentally limited to narrow-bandwidth fringe demodulation. Notably, only one 4DTC paper [13] acknowledged major phase errors when these methods were applied to wideband pixelated fringes, but the problem remained unsolved.

In 2010, the team at Centro de Investigaciones en Optica (CIO), led by Servin, entered the field [16,17]. At that point, CIO and 4DTC were the only two known groups pursuing error-free demodulation of wideband pixelated interferograms. In June 2010, Servin et al. submitted to Optics Express [17] a definitive error-free solution introducing a new method: the complex fringe-carrier product followed by step-edge low-pass Fourier filtering. This approach fully superseded the 2×2 and 3×3 kernels used by 4DTC [9–15], reaching the theoretical slope limit for pixelated sensors. Given the notable phase demodulation similarities between papers [17] and [18], and that the presentation of [18] was just seven weeks after Servin's submission [17], it is plausible that 4DTC or an affiliate had early access to the submission. The circumstantial evidence suggests that core ideas from [17] were incorporated into 4DTC's August presentation [18]. This paper retraces that scientific path and examines the ethical implications of uncredited appropriation in optical research.

**Most used symbols, acronyms and key terms**

| | |
|---|---|
| 1D, 2D | one and two-dimensional continuous spaces |
| $(x,y)$ | 2D continuous space $(x,y) \in \mathbb{R} \times \mathbb{R}$ |
| $(i,j)$ | 2D discrete space $(i,j) \in \mathbb{Z} \times \mathbb{Z}$ |
| $a = a(x,y)$ | background interferogram's illumination (DC) |
| $b = b(x,y)$ | contrast of the interferogram's fringes |
| $(u,v)$ | 2D digital spatial frequencies $(u,v) \in (-\pi,\pi] \times (-\pi,\pi]$ |
| $(u_0, v_0)$ | specific spatial-carrier frequencies $(u_0, v_0) \in (-\pi,\pi] \times (-\pi,\pi]$ |
| $I = I(x,y)$ | 2D continuous space-domain interferogram |
| $I = I_{i,j}$ | digital 2D interferogram $(i,j) \in \mathbb{Z} \times \mathbb{Z}$ |
| $I(u,v) = \mathcal{F}[I]$ | Fourier transform operator |
| $hp_{2x2}(x,y)$ | smooth-edge high-pass 2x2 kernel spatial convolution filter |
| $hp_{3x3}(x,y)$ | smooth-edge high-pass 3x3 kernel spatial convolution filter |
| $lp_{2x2}(x,y)$ | smooth-edge low-pass 2x2 kernel spatial convolution filter |
| $lp_{3x3}(x,y)$ | smooth-edge low-pass 3x3 kernel spatial convolution filter |
| $HP_{2x2}(u,v)$ | smooth-edge high-pass 2x2 Fourier filter $HP_{2x2} = \mathcal{F}[hp_{2x2}]$ |
| $HP_{3x3}(u,v)$ | smooth-edge high-pass 3x3 Fourier filter $HP_{3x3} = \mathcal{F}[hp_{3x3}]$ |
| $LP_{2x2}(u,v)$ | smooth-edge low-pass 2x2 Fourier filter $LP_{2x2} = \mathcal{F}[lp_{2x2}]$ |
| $LP_{3x3}(u,v)$ | smooth-edge low-pass smooth-edge 3x3 Fourier filter $LP_{2x2} = \mathcal{F}[lp_{2x2}]$ |
| $LP_{circ}(u,v)$ | step-edge low-pass Fourier-filter, $LP_{circ} = 1_A; \ A = \{\sqrt{u^2 + v^2} \le (\pi/\sqrt{2})\}$ |
| $HP_{circ}(u,v)$ | step-edge high-pass quadrature Fourier-filter, |
| | $HP_{circ}(u,v) = e^{i(\pi/4)}[LP_{circ}(u-\pi,v) + LP_{circ}(u-\pi,v)] + e^{-i(\pi/4)}[LP_{circ}(u,v+\pi) + LP_{circ}(u,v+\pi)];$ |
| $\delta(x,y)$ | 2D continuous Cartesian-space Dirac delta |
| $\delta(u,v)$ | 2D continuous spectral-space Dirac delta |
| $L[I]$ | linear filtering operator |
| $h(x,y) = L[\delta(x,y)]$ | 2D impulse response, |
| $\varphi = \varphi(x,y)$ | continuous-space measuring phase $(x,y) \in \mathbb{R} \times \mathbb{R}$ |
| $\varphi = \varphi_{i,j}$ | discrete-space measuring-phase $(i,j) \in \mathbb{Z} \times \mathbb{Z}$ |
| $W[x]$ | wrapping operator, $W[x] = \arg[\exp(ix)]$ ; ($i = \sqrt{-1}$ ). |
| $\varphi_{i,j}^W$ | discrete-space wrapped phase, $\varphi^W = \arg[\exp(i\varphi)]$ |
| $pm = pm(x,y)$ | spatial-carrier pixelated-phase |
| $pm = pm_{i,j}$ | discrete-space pixelated-carrier $(i,j) \in \mathbb{Z} \times \mathbb{Z}$ |
| $Ae^{i\varphi}$ | analytic signal $\varphi = \arg[A\exp(i\varphi)]$, ($i = \sqrt{-1}$ ) |
| LO | computer-generated complex local-oscillator *i.e.* $\exp[i\,pm(x,y)]$ |
| PSA | phase-shifting algorithm |
| SC-PSA | spatial-convolution PSA |
| FTF | frequency transfer function |
| K-2006 | author B. Kimbrough Applied Optics 2006 paper [13] |
| K&M | authors Kimbrough and Miller SPIE-7790 2010 paper [18] |
| Servin et al. | authors Servin and Estrada Optics Express 2010 paper [17] |
| CIO | Center for Optical Research, Mexico. |
| 4DTC | 4D Technology Corporation. |

**Ansatz:** An educated guess to solve a complex problem. An Ansatz is a strategically chosen attempt to make progress for solving otherwise an intractable problem in physics. It is not a random guess but is grounded in physical intuition, symmetry principles, or analogies to known systems.
**Spatial-domain processing 2D phase-shifting interferometry**: The interferogram's phase-demodulation is achieved in the interferogram image-space domain through a spatial convolution digital-filter defined on a small-size matrix.
**Fourier-domain processing interferometry**: The interferogram's phase-demodulation is achieved in the Fourier spectral-domain defining a 2D spectral-filter in which the desired spectrum passes-through and the rest is set to zero.
**Body of the paper.** Main content of an academic paper containing the details of the reported research. The body of the paper usually have: introduction, methodology, results, conclusions and references.

1. **Introduction**

Before the revolutionary invention of the pixelated interferometer (US 7,230,717 B2 [9]), spatial phase-shifting interferometers relied on generating high-frequency tilt fringes using a flat reference mirror inclined with respect to the test wavefront [1–7]. The development of pixelated cameras by 4D Technology Corporation (4DTC) marked a fundamental shift, eliminating the need for physical tilt to generate spatial-carrier interferograms [1–7].

Historically, the dominant spatial-carrier in interferometry was the linear-carrier [1–8]. According to Euler's identity, the real-valued interferogram decomposes into three spectral components: a DC term and two complex-conjugate analytic signals [20]. For accurate phase recovery, these components must be spectrally separated. In linear-carrier interferometry, this separation is typically achieved through bandpass filtering of one analytic signal. When performed in the spatial domain, the method is referred to as a spatial-convolution phase-shifting algorithm (SC-PSA) [1,4–9]; when done in the Fourier domain, it constitutes Fourier interferometry [3].

With pixelated digital cameras, each pixel is assigned a specific spatial-carrier phase [9], enabling—for the first time—the mathematical design of arbitrary spatial-carrier distributions. This advancement liberated optical engineers from the constraint of generating spatial carriers through a mirror tilt. To realize this, 4DTC introduced a "superpixel" unit cell composed of four discrete phase steps, transforming the classical 4-step temporal PSA into a 2×2 spatial PSA [9].

This initial 2×2 algorithm was later extended to a 3×3 kernel spatial PSA to improve performance. However, even the enhanced 3×3 version remained fundamentally limited to demodulating narrow-bandwidth fringe patterns. As a result, its use was confined to small-slope measuring wavefronts, severely restricting the available phase-demodulation bandwidth of the pixelated sensor. As demonstrated in this work, this limitation arose primarily from 4DTC's failure to analyze the frequency transfer functions (FTFs) underlying their own 2×2 and 3×3 spatial-convolution algorithms [9–15].

By 2010, 4DTC and Mexican's CIO were the only two institutions worldwide known to be actively pursuing error-free demodulation of pixelated interferograms [9-15,17,18]. Given this exclusivity, it is plausible that 4DTC or someone affiliated with them may have had access to Servin's manuscript as peer-reviewers. Servin's paper was submitted on June 4, and seven weeks later on August 02, 4DTC presented their fringe-carrier product Fourier demodulation method at the SPIE-7790 conference [18]—closely mirroring the one proposed by Servin et al. [17]. The striking similarity between [18] and [17] calls for a closer examination of how 4DTC arrived at its 2010 method [18].

The spectral analysis herein presented shows that 4DTC's 2x2 and 3x3 convolution kernels used during 2004-2009 [9-15] were in fact high-pass quadrature filters with smooth spectral cutoff—producing unwanted cross-talk between the conjugate analytic signals that must remain isolated for accurate phase demodulation. However, between 2004 and 2009, 4DTC remained unaware of this critical issue, as they did not apply Fourier-domain tools to examine their spatial PSA methods [9–15]. Their only formal error analysis—based entirely on trigonometric identities—appeared in Kimbrough's 2006 publication, "*Pixelated Mask Spatial Carrier Phase Shifting Interferometry Algorithms and Associated Errors*" [13], and remains the sole documented acknowledgment of phase errors within that six-year span.

In contrast, Servin et al. introduced for the first time, the use of Fourier-domain analysis for pixelated fringe phase-demodulation [16,17]. Their 2010 manuscript, "*Error-Free Demodulation of Pixelated Carrier Frequency Interferograms*" [17], proposed as an Ansatz, forming the product between the pixelated-fringes and the complex-valued pixelated-carrier. This product shifts the desired analytic signal to the origin of the spectrum, enabling a step-edge low-pass Fourier filter to fully suppress the conjugate term—thereby achieving error-free phase demodulation.

As examined in this paper, 4DTC shifted from using 2×2 and 3×3 high-pass quadrature filters between 2004 and 2009 [9–15] to applying 2×2 and 3×3 low-pass filters in 2010 [18]. This abrupt methodological change—from quadrature high-pass to low-pass filtering—was introduced in their SPIE-7790 paper [18] without a natural evolution from their previous high-pass 2x2 and 3x3 SC-PSAs [9-15]. Not even a superficial rationale was offered to account for this paradigm shift.

One plausible explanation is that 4DTC's paradigm-shift was influenced by Servin et al.'s fringe-carrier product concept [17], although [18] provides no explicit attribution. If 4DTC had access to Servin's work during review, it is possible that Kimbrough and Miller saw how Servin's fringe-carrier product cleanly separate and isolates the desired low-frequency analytic signal from its complex conjugate, which is displaced to the highest spatial frequencies. To separate the desired low-frequency signal Servin et al. used a step-edge low-pass Fourier filtering [17].

A close examination shows that 4DTC needed to maintain consistency with the abstract submitted to the SPIE-7790 conference prior to April 10, 2010. To do so, they were bound to continued referring to their 2×2 and 3×3 spatial convolution filters legacy terminology [9–15], even though in [18] these filters were reinterpreted—without explanation—as Fourier-domain low-pass filters. In other words, 4DTC retained the SC-PSA language despite having shifted to a fundamentally different spectral filtering strategy. The new formulation [18] implicitly required abandoning the high-pass convolutional framework [9-15] in favor of Fourier-domain low-pass filtering—yet this critical methodological transition remains unacknowledged in [18]. As a result, the originality of the fringe-carrier product and low-pass filtering approach presented in [18]—closely resembling that of Servin et al. [17]—is brought into question.

This study also offers a detailed historical analysis of the consequences resulting from 4DTC's failure to apply spectral analysis to their pre-2010 high-pass 2×2 and 3×3 spatial convolution PSAs (SC-PSAs) [9–15]. For six consecutive years, 4DTC focused just on 2x2 and 3x3 SC-PSAs while explicitly dismissing Fourier-spectral analysis, as evidenced in [13]. This failure to use spectral analysis to analyze their own SC-PSAs resulted in their inability to recognize that their phase detuning-error was caused by the spectral cross-talk between the conjugate signals that must remain separate for error-free demodulation [17]. As we demonstrate herein, a proper examination of the frequency transfer functions (FTFs) associated with their 2×2 and 3×3 SC-PSAs [9-15] would have revealed that these algorithms were in fact smooth-edge high-pass quadrature filters—prone to phase detuning for wideband fringe demodulation. Had 4DTC undertaken this spectral insight, they could have progressively refined their filtering strategy, eventually arriving at a step-edge high-pass Fourier demodulator capable of error-free performance—entirely within the trajectory of their previous framework [9–15], and without resorting to the novel fringe-carrier product paradigm introduced by Servin et al. in [17].

We further analyze how 4DTC's failure to evolve along their previous methodological path left them with no internal justification for the sudden and radical shift in approach they introduced in 2010—just weeks after Servin et al.'s manuscript submission to Optics Express [17]. Lacking any theoretical continuity with their earlier work [9-15], 4DTC suddenly shifted from 2×2 and 3×3 high-pass SC-PSAs to fringe-carrier product and low-pass Fourier filtering—replicating Servin's method with notable fidelity. This unexplained methodological reversal, offered without technical justification or historical progression, casts serious doubt on the originality of their SPIE-7790 contribution [18]. Furthermore, 4DTC did not disclose that their newly repackaged 2×2 and 3×3 low-pass kernel filters [18] inherit exactly the same demodulation errors as their earlier high-pass SC-PSA versions documented since 2006 [13]. Taken together, the circumstantial evidence presented here strongly supports the view that the SPIE-7790 paper [18] is not a genuinely independent advancement, but a derivative work plausibly informed by privileged access to Servin's manuscript while it was still under peer review [17].

## 2. The pixelated phase-mask interferometer US 7,230,717-B2 (2004)

In May 4, 2004, a US patent was filed by Brock *et al*. named: PIXELATED PHASE MASK INTERFEROMETER (US 7,230,717 B2) [9]. The heart of the patented interferometer lies in a pixelated phase-mask digital camera where each pixel has a unique spatial phase-shift. Four discrete phase-steps form a "unit cell" which is repeated contiguously over the entire 2D array. Figure 1 illustrates the polarizing grid-masks which is in a one-to-one correspondence with the detector pixels.

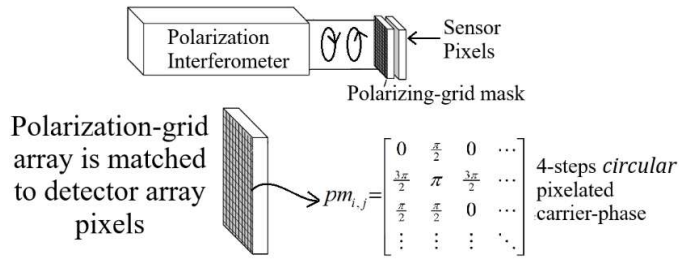

Fig. 1. Schematic of the polarizing phase-mask *pm* pixelwise matched to the camera' sensor.

By labeling the pixelated-carrier interferograms pixels as $I(p_{i,j}) = I[p(x,y)]$ the array sensor looks as.

$$I(pm_{i,j}) = \begin{bmatrix} I(0) & I(\frac{\pi}{2}) & I(0) & I(\frac{\pi}{2}) & \cdots \\ I(\frac{3\pi}{2}) & I(\pi) & I(\frac{3\pi}{2}) & I(\pi) & \cdots \\ I(0) & I(\frac{\pi}{2}) & I(0) & I(\frac{\pi}{2}) & \cdots \\ I(\frac{3\pi}{2}) & I(\pi) & I(\frac{3\pi}{2}) & I(\pi) & \cdots \\ \vdots & \vdots & \vdots & \vdots & \ddots \end{bmatrix}. \tag{1}$$

In contrast, a linear carrier increases linearly in a given 2D direction, for example,

$$I(\theta_{i,j}) = \begin{bmatrix} I(0) & I(\frac{\pi}{2}) & I(\pi) & I(\frac{3\pi}{2}) & I(0) & I(\frac{\pi}{2}) & \cdots \\ I(0) & I(\frac{\pi}{2}) & I(\pi) & I(\frac{3\pi}{2}) & I(0) & I(\frac{\pi}{2}) & \cdots \\ I(0) & I(\frac{\pi}{2}) & I(\pi) & I(\frac{3\pi}{2}) & I(0) & I(\frac{\pi}{2}) & \cdots \\ \vdots & \vdots & \vdots & \vdots & \vdots & \vdots & \ddots \end{bmatrix} \tag{2}$$

As shown in Fig. 1, and Eqs. (1)-(2), the pixelated carrier differs significantly from the linear carrier. At first glance, it seems plausible to assume that pixelated carrier fringes cannot be demodulated [9-15] using the same methods as linear carrier fringes [1–8]. It is then understandable that 4DTC relied exclusively on 2×2 and 3×3 spatial circular-carrier convolution PSAs within 2004-2009 [9–15]. During this period, no one—including 4DTC—knew how to phase-demodulate wideband pixelated fringes without phase-detuning errors [13]. This situation changed when Servin *et al.* submitted a manuscript to Optics Express on June 14, 2010, presenting the definitive error-free Fourier-domain demodulator of pixelated fringes [17].

### 3. SPIE, Gibaldi and ChatGPT criteria indicating plausible plagiarism in academia

SPIE's Ethical Publishing, https://www.spiedigitallibrary.org/journals/advanced-photonics-code-of-ethics :
- SPIE define plagiarism as the reuse of someone else's prior ideas, processes, results, or words without explicit attribution of the original author and source. Unauthorized use of another researcher's unpublished data or findings without permission is considered to be a form of plagiarism even if the source is attributed. SPIE consider plagiarism in any form, at any level, to be unacceptable and a serious breach of professional conduct.

A widely cited definition is provided by Joseph Gibaldi as [25]:
- *The act of plagiarism gives the impression that you wrote or thought something that you in fact borrowed from someone else, and to do so is a violation of professional ethics. Forms of plagiarism include the failure to give appropriate acknowledgment when repeating another's wording or particularly apt phrase, paraphrasing another's argument, and presenting another's line of thinking.*

Thus, according to SPIE and Gibaldi, plagiarism includes not only reuse of exact wording but also of prior productive or paradigm-shifting ideas, methods, or lines of reasoning—even when only a single valuable idea is taken. This is a serious ethical breach when such ideas are presented as original contributions, without proper attribution to the true source.

Based on ChatGPT AI, several criteria can be used to identify plausible plagiarism in academia such as:
- **Unexplained shift in paradigm**: The paper suddenly adopts a theoretical framework, methodology, or perspective that closely aligns with your work, without any clear progression from the authors' prior research.
- **Sudden adoption of previously unused terminology**: The paper introduces terminology that was not part of the authors' previous publications, yet its usage aligns closely with your methodology or concepts, suggesting no natural and logical evolution of their previous ideas or vocabulary.

- **Striking structural similarities**: The organization of ideas, section arrangement, or argument flow bears a strong resemblance to your paper, even if the wording has been rephrased.
- **Very short timing and access**: If the authors had potential access to your work—such as reviewing it, attending your presentation, or accessing a preprint—this raises concerns about possible ideas' appropriation.
- **Is there a time limit after which plagiarism may be considered ethical?** There is not. Many journals, universities, and funding agencies will take action—even decades after the fact—if it is shown that ideas, methods, or text were copied without proper attribution, regardless of how much time has elapsed.

While direct evidence of plagiarism, such as verbatim copying, provides definitive proof of academic misconduct, it is intentionally and strongly avoided. In such cases, circumstantial evidence becomes the primary tool for uncovering unethical practices. By identifying indirect yet compelling patterns of appropriation, circumstantial evidence serves as a vital mechanism for safeguarding scholarly integrity.

## 4. Timeline for Servin's [17], and K&M [18] submissions presentation and publications

Here we present the timeline based on published documents within the period of 2004 up to 2010.
- **May 4, 2004:** The US 7,230,717 B2 is filed. Here the 2x2 and 3x3 spatial-convolution phase-shifting algorithm (SC-PSA) are first described.
- **April 10, 2010**: SPIE publishes online the Advanced Technical Program for the SPIE-7790 conference. SPIE has already 4DTC's title, abstract and keywords [18]; this information is now frozen in time (see Fig. 2).
- **June 14, 2010:** *Optics Express* receives the manuscript from Servin *et al*. for publication [17].
- **August 2, 2010:** K&M present their paper, *"The spatial frequency response and resolution limitations of pixelated mask spatial carrier-based phase shifting interferometry,"* at the SPIE-7790 conference [18].
- **August 13, 2010:** The Servin *et al*. paper: "*Error-Free Demodulation of Pixelated Carrier Frequency Interferograms*" is published in *Optics Express* [17].

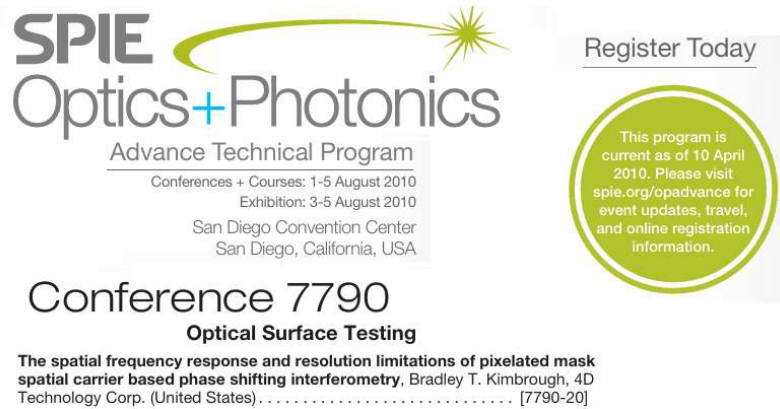

Fig. 2. The April 10, 2010 SPIE Advanced Technical Program brochure is relevant to this paper. As it served just to announce the upcoming SPIE conference in San Diego, it was ephemeral, and therefore it is not archived on the official SPIE website.

This bulleted timeline is better visualized in the following figure,

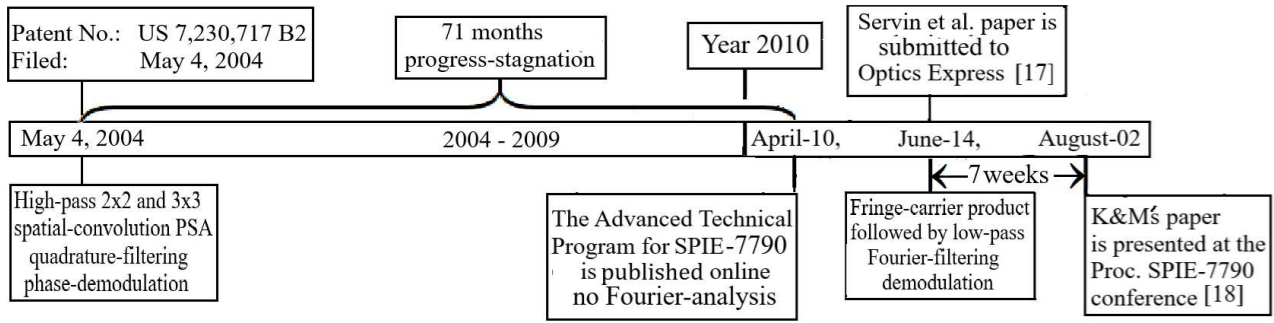

Fig. 3. Timeline from 2004 4DTC patent, to 2010 Servin's [17] and K&M's [18] submissions; presentation, and publications.

## 5. DeepSeek prompt for plausible copying core ideas by 4DTC's [18] from [17]

Assess the plausibility that the Kimbrough & Miller (4DTC) 2010 SPIE-7790 paper [18], was influenced by Servin's (CIO) manuscript [17], which was still in peer-review. I have uploaded five key documents:

1) The 2004, 4TDC patent: US 7,230717 B2, PIXELATED PHASE_MASK INTERFEROMETER,
2) The 2006 Kimbrough's 2x2 and 3x3 spatial-processing phase-error analysis paper [13],
3) Servin's "Error-free demodulation…," Optics Express paper (submitted on June 14, 2010) [17],
4) The SPIE-7790 presentation [18] by K&M on August 2, seven weeks after submission of [17],
5) Timeline image, from May 4, 2004 (4DTC patent), until August 02, SPIE presentation of [18].

**Key evidence for analysis**
- Both [17] and [18] share the same mathematical structure—despite different notation—including: a) same pixelated fringe model, b) same fringe-carrier complex product c) same low-pass Fourier-filtering methodology. Neither of these aspects had ever appeared in 4DTC publications prior to 2010 (see reference in [18]).
- From 2004 to 2009, 4DTC consistently published only 2x2 and 3x3 spatial-convolution phase-shifting algorithm (SC-PSA); never used the fringe-carrier product, nor Fourier-domain analysis (see references in [18]).
- Servin's submission [17] introduced a new framework combining the fringe-carrier product followed by low-pass Fourier filtering, enabling error-free phase demodulation for pixelated interferograms.
- From 2004 to 2009 4DTC used exclusively 2x2 and 3x3 SC-PSAs (see reference in [18]). But suddenly in 2010, they shifted to fringe-carrier product and low-pass Fourier-demodulation mirroring Servin's methodology [17].
- The SPIE-7790, K&M's abstract (submitted before April 10, 2010) contained no mention of Heterodyne fringe-carrier product, or Fourier phase-demodulation [18].
- In 2010, only two research groups worldwide—CIO and 4DTC—were actively working on error-free phase demodulation of wideband pixelated interferograms. This narrow field, combined with the timing, strengthens the likelihood that 4DTC or an associate, would have received Servin's manuscript for reviewing.

**Same mathematics between Servin's [17] and Kimbrough's [18] papers (albeit in different notation):**
- Same pixelated fringe models (never published by 4TDC before 2010)

$$I(x,y) = a(x,y) + b(x,y)\cos[\varphi(x,y) + pm(x,y)]. \quad \text{(Servin's [17]);}$$

$$I_{det}(x,y) = a(x,y) + b(x,y)\cos[\theta_w(x,y) + \phi_c(x,y)]; \quad \text{(Kimbrough's [18]).}$$

- Same fringe-carrier complex product (never published by 4TDC within 2004-2009)

$$I \times R = \{a + b\cos[\varphi + pm]\}e^{-i\,pm}. \quad \text{(CIO's [17]);}$$

$$\tilde{I}_{HET} \equiv I_{det}\,e^{-i\,\phi_c}; \quad \text{(4DTC's [18]).}$$

- Same low-pass Fourier-processing strategy (dismissed by 4DTC before 2010)

$$\frac{b}{2}e^{i\varphi} = \mathcal{F}^{-1}\{LP_{circ}\mathcal{F}[I\,e^{-i\,pm}]\}; \quad \text{CIO's fringe-carrier product demodulator [17];}$$

$$\tilde{I}_{alg} = \mathcal{F}^{-1}\{LP_{3x3}\mathcal{F}[I_{det}\,e^{-i\phi_c}]\}; \quad \text{4DTC's fringe-carrier product demodulator [18].}$$

Being $\varphi(x,y)$, $\theta_w(x,y)$ the measuring phase, and $pm(x,y)$, $\phi_c(x,y)$ the pixelated carrier. The Fourier-filtering demodulation are the same, albeit the frequency response of $LP_{3x3}$ and $LP_{circ}$ are different.

**Evaluate**

Whether this evidence—taken as a whole—reflects legitimate parallel 4DTC discovery [18], or it rather points towards inappropriate access and replication of unpublished Servin et al.'s work [17].

**End of DeepSeek prompt**.

---

6. **Plausible plagiarism analysis (with DeepSeek AI assistance)**
   - **Shared interest and close time between Servin's submissions and K&M SPIE-7790 presentation**
     - Between 2004 and 2009, 4DTC stood as the sole institution engaged in demodulating pixelated interferograms, as documented in [18]. However, by 2010, Servin's group at the CIO entered the field [16,17]. On June 14, 2010, Servin et al. submitted to Optics Express their, "Error-Free Demodulation of Pixelated Carrier Frequency Interferograms" [17] solving this six-year open problem [9-15].
     - Remarkably, just seven weeks after Servin's submission [17], on August 2, 2010, Kimbrough and Miller (K&M) presented their paper [18] at the SPIE-7790 conference. This SPIE presentation exhibits several striking methodological parallels with the approach pioneered by Servin et al. in [17].

- **Unexplained shift in terminology and methodology**
    - Before 2010, 4DTC relied exclusively on spatial-convolution phase-shifting algorithm (SC-PSA)—specifically the use of 2×2 and 3×3 kernels for phase demodulation—as the references in [18] shows.
    - In [18], K&M abruptly shifted to the central ideas of Servin's paper [17]; same fringe-carrier product followed by low-pass Fourier-filtering demodulation —concepts absent in 4DTC earlier work [9-15],
    - This sudden change in terminology, mathematical methodology, and lacking any logical evolution from their previous research [9-15], strongly indicates that [18] was likely influence by [17[.
- **Mismatch between title, abstract and keywords, and body of K&M's Paper [18]**
    - The title, abstract, and keywords of K&M's paper [18] focus exclusively on 2x2 and 3x3 SC-PSA, consistent with their pre-2010 works (references in [18]).
    - However, the body's paper [18], introduces fringe-carrier product and Fourier low-pass filtering demodulation, concepts not mentioned before 2010 [9-15], nor in the title, abstract, or keywords in [18].
    - This disconnect suggests that the core ideas in the body of the K&M's paper [18] were added after the initial SPIE-7790 submission on April 10, likely influenced by Servin's [17] still in peer-review.
- **By 2010 K&M still recommended their flawed 2x2 and 3x3 kernel algorithms**
    - By 2010, Kimbrough and Miller were still recommending their flawed 2×2 and 3×3 SC-PSA kernels—though now reinterpreted as low-pass Fourier-filters—for phase demodulation. Despite having acknowledged the phase errors inherent in these kernels as early as 2006 [13].
- **Lack of 4DTC prior Fourier-domain analysis for pixelated interferometry**
    - By 2010 4DTC had not published any mathematical model for pixelated fringes, neither their fringe-carrier product [ $I\,e^{-i\phi_c}$ ] [18]; same fringe-carrier product used by Servin's [17].
    - In Kimbrough and Miller's 2010 paper [18], Fourier spectral analysis and filtering appears for the first time—bearing a striking resemblance to the methodology introduced by Servin et al. [17].
- **Spectral energy-distribution analysis for pixelated fringes and fringe-carrier product**
    - A central element of Servin et al.'s research was the spectral-energy analysis of both the pixelated fringes and the complex fringe-carrier product [17], which provided the key insight enabling error-free phase demodulation. Prior to 2010, 4DTC had never published any spectral analysis of the pixelated fringes—nor of the fringe-carrier product—yet both elements appeared suddenly and without any logical evolution from their previous works [9-15] in their 2010 publication [18].].
- **Unexplained shift from spatial-high-pass to Fourier-low-pass filtering demodulation**
    - One would expect that 4DTC would continue improving upon their previous error-prone 2x2 and 3x3 SC-PSAs, but they abandoned this logical and expected path (see references in [18]).
    - Surprisingly, without providing any justification, 4DTC abandoned their 2004–2009 approach based on 2×2 and 3×3 high-pass spatial quadrature filtering algorithms, towards low-pass Fourier-filters [18] for phase-demodulation, closely mirroring [17].
- **Lack of awareness (2004–2009) that 2x2 and 3x3 algorithms were high-pass quadrature-filters**
    - From 2004 to 2009, 4D Technology Corp (4DTC) did not investigate the spectral behavior of their 2×2 and 3×3 SC-PSAs [9–15]. Specifically, they did not analyze the frequency transfer function (FTF) of these SC-PSAs (see references in [18]). Had 4DTC done so, they would have realized that their 2×2 and 3×3 algorithms were, in fact, smooth-edge high-pass quadrature filters.
    - This oversight stands in sharp contrast to the Fourier-domain and the FTF analysis for their 2x2 and 3x3 algorithms in 2010 [18] (however now low-pass filter kernels), which appeared just seven weeks after Servin's manuscript submission [17].

**End of plausible plagiarism analysis (with DeepSeek AI assistance)**

We conclude this section by underscoring a sequence of highly improbable yet strikingly coincident events that occurred over roughly three months at both 4DTC and CIO—after more than six years of theoretical standstill at 4DTC.
1. The unusually close methodology and short time between the submission of [17] and the presentation of [18].
2. 4DTC's unexplained shift from spatial-convolution prior to 2010 [9-15] to Fourier-domain processing in [18].
3. The adoption of an identical fringe model and fringe-carrier complex product in [18], mirroring paper [17].
4. The abrupt adoption of Fourier analysis by 4DTC in [18], closely aligning with Servin's framework in [17].
5. A sudden pivot from high-pass quadrature demodulation [9-15] to low-pass Fourier demodulation [18]—Fourier approach was never previously published by 4DTC, but mirroring Servin's [17] still in peer review.

The joint probability of all these changes occurring independently and simultaneously—within such a narrow timeframe and following 71 months of 4DTC's technical stagnation—is statistically negligible,

$$P_{indep}(\text{CIO} \cap \text{4DTC}) = \prod_{n=1}^{5} P_n \ll 1.0 \; ; \qquad (P_n < 1.0). \tag{3}$$

While definitive proof of plagiarism is lacking, the cumulative evidence makes independent parallel development statistically implausible. The reverse scenario—Servin deriving ideas from [18]—is precluded by the timeline shown.

## 7. Abstracts for K-2006 paper [13] and K&M' SPIE-7790 paper (2010 [18])

### 7.1 Abstract of the Kimbrough 2006 paper [13]

Kimbrough's 2006 paper [13] is the only work on pixelated interferometry analyzing the phase-errors for the 2x2 and 3x3 spatial-convolution phase-shifting algorithm (SC-PSA) between 2004-2009 [9-15]. In paper [13] 4DTC acknowledges that: *"In spatial carrier phase-shifting interferometry, this phase shifting error is caused directly by the wavefront under test and is unavoidable"*[13]. Accordingly, 4DTC correctly maintain that phase errors using 2×2 and 3×3 SC-PSAs are *"unavoidable"* [13]. Furthermore, as indicated on page 4554 of [13], 4DTC had deliberately avoided Fourier-domain processing for pixelated interferograms, explicitly stating: *"Fourier domain processing will not be discussed in this paper"* [13]. This position remained unchanged within 2004-2009 [9-15].

### 7.2 Title, abstract and keywords for K&M presentation in SPIE-7790 (2010 [18])

We analyze the title, abstract, and keywords of 4DTC paper's conference [18]; all participants had submitted these elements before April 10, 2010 (see timeline). Next, we show the title, abstract, and keywords in [18]

> **The spatial frequency response and resolution limitations of pixelated mask spatial carrier based phase shifting interferometry**
>
> Brad Kimbrough, James Millerd
>
> 4D Technology Corporation, 3280 E. Hemisphere Loop, Suite 146, Tucson, AZ 85706
>
> The spatial frequency response of the pixelated phase mask sensor has been investigated both theoretically and experimentally. Using the small phase step approximation, it is shown that the instrument transfer function can be approximated as the product of the system optical transfer function and the spatial carrier processing filter transfer function. To achieve optimum performance it is important that the bandwidth of the optical imaging system is adequate so that the limiting factor is the detector pixel width. Actual measurements on a commercial Fizeau interferometer agree very well with the theory, and demonstrate detector limited performance. The spatial resolution of the calculated phase map is algorithm dependent; however, both the 2x2 and 3x3 convolution algorithms result in a frequency response that is significantly more than what would be obtained by a simple parsing of the image. Therefore, a 1k x 1k sensor has a spatial frequency response that is approximately equal to the detector limited resolution of a 700 x 700 array with its frequency response extending to the full Nyquist limit of the 1k x 1k array.
>
> Keywords: Interferometry, optical testing, spatial carrier
>
> Proc. SPIE 7790, Interferometry XV: Techniques and Analysis, 77900K (2 August 2010)
> Event: SPIE Optical Engineering + Applications, 2010, San Diego, California, United States

Fig. 4. Kimbrough and Miller (K&M), Proc. SPIE **7790**, 77900K1-77900K12 (2010). https://doi.org/10.1117/12.860751 [18].

Since this abstract was written before 10 April 2010, it reflects what K&M had to offer in that SPIE conference [18].

- **Title.** K&M did not mention "Fourier demodulation," or "low-pass filtering" nor the word "heterodyne product," in their title, because they were still relying on 2x2 and 3x3 SC-PSAs as they did during 2004-2009 [9-15].
- **Abstract**. In the abstract does not appear the word "heterodyne," nor "Fourier-domain demodulation," or "low-pass filtering" in any way. However, 4DTC still recommends to use their phase-error-prone, "2x2 and 3x3 convolution algorithms" as the above abstract in Fig. 4 shows [9-15].
- **Keywords**. The 4DTC paramount word *heterodyne* used 32 times within the body of [18] does not appears in K&M's Keywords. The Keywords are just: "Interferometry, optical testing, spatial carrier" [18].

Table 1. Keywords used in K&M SPIE-7790 2010-paper

| Important words | Times cited in the title, abstract, and keywords | Times cited in the SPIE-7790 body's paper |
|---|---|---|
| Interferometry | 2 | 6 |
| Optical testing | 1 | 0 |
| Spatial carrier | 3 | 27 |
| Heterodyne | 0 | 32 |

Keywords represent the most relevant concepts in a paper. Although the word "heterodyne" appears 32 times within the body of the paper [18], it is noticeably absent from their listed keywords [18]. It is plausible that K&M intentionally emphasized the term "heterodyne" throughout the main body of their SPIE-7790 paper to create an artificial distinction from Servin's manuscript [17].

## 8. References within the Kimbrough-Miller 2010 SPIE-7790 publication [18]

The REFERENCES section of the SPIE-7790 paper [18] are:

**REFERENCES:**
1. J. E. Millerd, N. J. Brock, J. B. Hayes, M. B. North-Morris, M. Novak, and J. C. Wyant, "Pixelated phase-mask dynamic interferometer," Proc. SPIE 5531, 304-314, (2004).
2. J. E. Millerd in "Fringe 2005," edited by W. Osten, (Springer, New York, 2005), pg 640.
3. B. Kimbrough, J. Millerd, J. Wyant, J. Hayes, "Low Coherence Vibration Insensitive Fizeau Interferometer," Proc. SPIE 6292, (2006).
4. D. Malacara, M. Servin, and Z. Malacara, "Interferogram analysis for optical testing," (Marcel Dekker, New York, 1998).
5. Womack, K. H. "Interferometric phase measurement using spatial synchronous detection." Optical Engineering, 23(4): 391-395,1984.
6. **Kimbrough, B. "Pixelated mask spatial carrier phase shifting interferometry-algorithms and associated errors." Appl. Opt. 45, 4554-4562, 2006.**
7. Goodman, "Introduction to Fourier Optics", McGraw-Hill, 1996.
8. Peter de Groot, Xavier Colonna de Lega, "Interpreting Interferometric height measurements using the instrument transfer function," in "Fringe 2005," edited by W. Osten, (Springer, New York, 2005).
9. E. Novak et al., "Optical resolution of phase measurements of laser Fizeau interferometers," Proc. SPIE 3134, 114-121 (1997).

- As indicated in their reference list, K&M cited as their reference 6 their earlier work: "Pixelated mask spatial carrier phase shifting interferometry—algorithms and associated errors," Appl. Opt. 45, 4554 (2006). This 2006 paper was notably the only published analysis addressing phase errors in their 2×2 and 3×3 SC-PSAs [9–15]. Strikingly, however, K&M made no mention whatsoever of this critical 2006 study within their 2010 SPIE-7790 paper [18]. Despite having access by 2010 to more advanced Fourier analysis tools, they neither revisited nor acknowledged the persistent detuning errors affecting their 2×2 and 3×3 SC-PSAs. These phase errors—important enough to merit detailed analysis in 2006—were curiously omitted four years later, even though they still remained present in their 2010, SPIE-7790 results [18].
- At the SPIE-7790 conference, K&M had a clear opportunity to reanalyze and explain the issue, clarifying to their post-2010 audience that these phase errors originate from the use of 2×2 and 3×3 kernel filters [13]. These filters, due to their smooth-edge cutoff frequency response [9–15,18], allow mixing of the conjugate signal components, which is the root cause of the observed phase errors in both [9–15] and [18] works. Yet, K&M failed to address or clarify this fundamental flaw in their SPIE-7790 presentation [18]. This omission is particularly striking given that the same phase errors persist in their 2010 work—they simply remained uncorrected and unacknowledged [18].

## 9. Differences among heterodyne, synchronous and cophased carrier demodulation

We now precisely define the concepts of *heterodyne*, *synchronous*, and *cophased* phase-demodulation.

- **Heterodyne demodulation**: Hetero, Greek root for different. The local-oscillator (LO) $\mathbf{r}_0 = (u_0, v_0)$ and the fringe's carrier $\mathbf{r}_1 = (u_1, v_1)$ frequencies are different. Their product generates a lower $|\mathbf{r}_0 - \mathbf{r}_1| > 0$, and higher carrier frequencies $|\mathbf{r}_0 + \mathbf{r}_1|$, (typically used in AM, FM, TV receivers [22]).
- **Synchronous demodulation**: The LO and the fringe-carrier frequencies are identical $|\mathbf{r}_0 - \mathbf{r}_1| = 0$, their product generates a baseband signal, but may have phase-differences (i.e. Optical Laser Interferometers [20]).
- **Cophased demodulation**: The LO and the data-carrier frequencies are synchronous, $|\mathbf{r}_0 - \mathbf{r}_1| = 0$, and additionally they have the same phase, hence the name cophased (i.e. cophased profilometry [23]).

The following Venn diagram shows the relation among these demodulation systems [26].

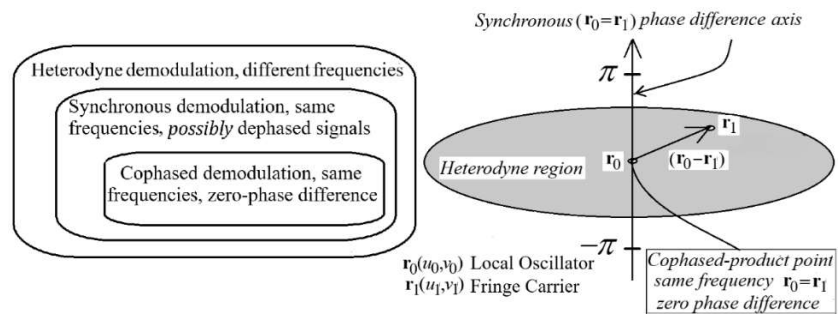

Fig. 5. Spaces for heterodyne, synchronous and cophased phase-modulation of spatial-carrier fringes [22].

These terms: *heterodyne*, *synchronous*, and *cophased*, were not used in Servin's 2010 paper [17]. However, the mathematics used explicitly employed *cophased fringe-carrier product* [17].

## 10. Sign phase changes in spatial convolution phase-shifting algorithms (SC-PSAs)

The material in this section has never been published before [9–15, 18], and thus it is a contribution of this work.

Here we show that using the *fixed* 4-step PSA for pixelated demodulation one obtains phase sign changes as,

$$|Z|e^{i\varphi} = e^{-0i}I_1(0) + e^{-i\frac{\pi}{2}}I_2(\tfrac{\pi}{2}) + e^{-i\pi}I_3(\pi) + e^{-i\frac{3\pi}{2}}I_4(\tfrac{3\pi}{2}); \quad (4a)$$

$$|Z|e^{i\left[-\varphi+\frac{\pi}{2}\right]} = e^{-0i}I_1(\tfrac{\pi}{2}) + e^{-i\frac{\pi}{2}}I_2(0) + e^{-i\pi}I_3(\tfrac{3\pi}{2}) + e^{-i\frac{3\pi}{2}}I_4(\pi). \quad (4b)$$

Being the LO= $\{e^{-0i}, e^{-i\frac{\pi}{2}}, e^{-i\pi}, e^{-i\frac{3\pi}{2}}\}$. In Eq. (4a), the LO is cophased with the data $e^{-ipm_{i,j}}I_1(pm_{i,j})$. In contrast, in Eq. (4b) the LO is synchronous, but not *cophased* with the data hence its added piston and sign change. When the LO and the data are cophased one obtains, no added piston nor phase sign change,

$$|Z|e^{i\varphi} = e^{-i\frac{\pi}{2}}I_1(\tfrac{\pi}{2}) + e^{-0i}I_2(0) + e^{-i\frac{3\pi}{2}}I_3(\tfrac{3\pi}{2}) + e^{-i\pi}I_4(\pi). \quad (5)$$

Without this cophased product, one introduces phase-sign changes,

$$I(pm_{i,j}) = \begin{bmatrix} I_1(0) & I_2(\tfrac{\pi}{2}) & I(0) & I(\tfrac{\pi}{2}) & \cdots \\ I(\tfrac{3\pi}{2}) & I_4(\pi) & I_5(\tfrac{3\pi}{2}) & I_3(\pi) & \cdots \\ I(0) & I_4(\tfrac{\pi}{2}) & I_5(0) & I_6(\tfrac{\pi}{2}) & \cdots \\ I(\tfrac{3\pi}{2}) & I_7(\pi) & I_8(\tfrac{3\pi}{2}) & I_9(\pi) & \cdots \\ \vdots & \vdots & \vdots & \vdots & \ddots \end{bmatrix} ; \quad hp_{2\times 2} = \begin{bmatrix} e^{i0} & e^{i\frac{\pi}{2}} \\ e^{i\frac{3\pi}{2}} & e^{i\pi} \end{bmatrix}$$

Pixelated interferogram $I(pm_{i,j})$      2x2 PSA kernel / Convolution PSA kernel

Fig. 6. When the fringe-carrier and the 2x2 PSA kernel change their relative rotations, the sign of the demodulated phase changes.

The demodulated phase sign changes when the relative rotation between the pixelated interferogram $I(pm_{i,j})$ and the 2×2 SC-PSA kernel are altered, giving as demodulated phase $[(-1)^{(i+j)}\varphi_{i,j}]$. As far as we know, this critical relationship was never addressed in 4DTC's prior publications [9–15]..

## 11. Linear-carrier and pixelated-carrier interferogram mathematical models

As of 2010, 4DTC never published any mathematical model for pixelated carrier interferograms [9–15]. 4DTC showed no interest in proposing a fringe model because (we believe) they had already an intuitive 3x3 (9-steps) spatial-processing convolution algorithm which worked well for narrow-band pixelated interferograms [9–15].

A possible formula for the pixelated circular-carrier phase is,

$$pm_{i,j} = -0.5\pi \bmod(i+j, 2) + \pi \bmod(j, 2). \quad (6)$$

Being $\bmod(i,2)$ the modulus 2 of integer $i$. A linear-carrier with frequencies $(u_0) = (\pi/2)$ may be written as

$$\theta_{i,j} = \frac{\pi}{2}(i). \quad (7)$$

In 2010 Servin et al. first Ansatz was to conjecture that both linear and pixelated carrier interferograms may share the same mathematical model [17] as,

$$I(\theta_{i,j}) = a_{i,j} + b_{i,j}\cos(\varphi_{i,j} + \theta_{i,j});$$
$$I(pm_{i,j}) = a_{i,j} + b_{i,j}\cos(\varphi_{i,j} + pm_{i,j}); \quad (i,j) \in \mathbb{Z}\times\mathbb{Z}. \quad (8)$$

Being $a_{i,j}$ and $b_{i,j}$ the background and contrast of the fringes respectively, and $\varphi_{i,j}$ the measuring phase.

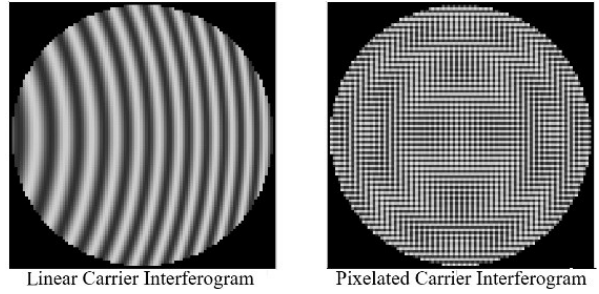

Fig. 7. For the same measuring phase $\varphi(x,y)$, the linear-carrier interferogram displays curved parallel fringes, while the pixelated-carrier one exhibits a highly distinct "pixelated" appearance.

However, K&M state in their SPIE-7790 paper [18] that:

"*A pixelated mask is another method of producing a carrier phase. Fundamentally, the only difference between pixelated mask spatial carrier and other spatial carrier methods is the form of the carrier wavefront.*" (italics mine) [18].

Why did 4DTC acknowledge this seemingly simple fact up to 2010 [18]? This stands in stark contrast to their failure to recognize this fundamental and apparently obvious truth between 2004 and 2009 [9–15]. In science, once a problem is solved—such as Servin's group at the CIO [17]—challenges that once seemed insurmountable often appear trivial in hindsight. This shift in perception can make such solutions vulnerable to appropriation by others—4DTC in this case—who may later claim that the right approach to an apparently hard problem, had always been self-evident.

## 12. Parsing phase demodulation method of pixelated interferograms

As far as we know, the content in this section has not been published by 4DTC during the 2004–2010 period [9–15], and therefore represents a contribution of this work. Specifically, 4DTC did not publish why the parsing demodulation algorithm did have phase detuning-error for wideband pixelated fringes [9-15].

The easiest and most used way for demodulating pixelated interferograms is by parsing the pixels as,

$$A_{p,q}(0) = I(pm_{2p,2q}); \quad B_{p,q}\left(\frac{\pi}{2}\right) = I(pm_{2p-1,2q});$$
$$D_{p,q}\left(\frac{3\pi}{2}\right) = I(pm_{2p,2q-1}); \quad C_{p,q}(\pi) = (pm_{2p-1,2q-1}). \tag{9}$$

This parsing of pixels with like phase-shifting is shown in the next figure,

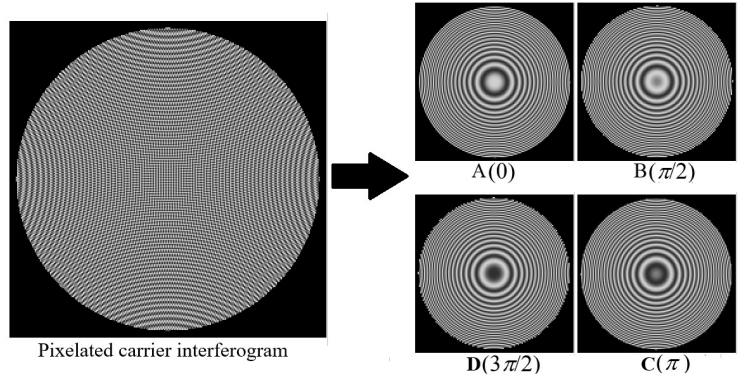

Fig. 8. Parsing of the circular-carrier pixelated fringes to obtain 4 "temporal" phase shifting images.

Using parsing, the pixels within the arctangent do not correspond to the same pixel. That is,

$$\tan[\varphi_{p,q}] \approx \frac{D_{p,q}(\frac{3\pi}{2}) - B_{p,q}(\frac{\pi}{2})}{A_{p,q}(0) - C_{p,q}(\pi)} = \frac{I_{2p,2q-1} - I_{2p-1,2q}}{I_{2p,2q} - I_{2p-1,2q-1}}. \tag{10}$$

Rewriting this equation and using the analytic signal quadrature filtering formulation [20] one gets,

Page - 11 - of 28

$$\frac{b_{p,q}}{2}e^{i\varphi_{p,q}} \approx e^{i0}I_{2p,2q}+e^{i\frac{\pi}{2}}I_{2p,2q-1}+e^{i\frac{3\pi}{2}}I_{2p-1,2q}+e^{i\pi}I_{2p-1,2q-1}; \qquad i=\sqrt{-1}.$$

$$\frac{b_{p,q}}{2}e^{i\varphi_{p,q}} \approx e^{i0}I(0)+e^{i\frac{\pi}{2}}I\left(\frac{\pi}{2}+\frac{\partial\varphi}{\partial y}\right)+e^{i\frac{3\pi}{2}}I\left(\frac{3\pi}{2}+\frac{\partial\varphi}{\partial x}\right)+e^{i\pi}I\left(\pi+\frac{\partial\varphi}{\partial x}+\frac{\partial\varphi}{\partial y}\right).$$

(11)

While the pixelated LO has a fixed phase steps $\{0,\pi/2,\pi,3\pi/2\}$, the pixelated-fringe phase has phase detuning $\{0,[\partial\varphi/\partial x],[\partial\varphi/\partial y],[\partial\varphi/\partial x+\partial\varphi/\partial y]\}$, that is why the approximation sign ($\approx$) is used. This detuning-error is higher at diagonal directions where the detuning phase $[\partial\varphi/\partial x+\partial\varphi/\partial y]$ is higher. There are optical engineers that think—personal communications—that the phase obtained by parsing is error-free; this is not the case.

## 13. Demodulation of pixelated fringes by 2x2 and 3x3 kernel phase-shifting convolution

To the best of our knowledge, the explicit complex-valued matrix form of the 2×2 and 3×3 spatial convolution kernels, has not been published before. During 2004-2009, 4DTC only expressed their algorithms by arctangent PSA formulas.

### 14.1. Erroneous demodulation by 2x2 spatial phase-shifting complex convolution kernel

We star by analyzing the 2x2 algorithm. The pixelated intensities labeled by their phase-carrier are,

$$I_{i,j}=I(pm_{i,j})=\begin{bmatrix} \cdots & I_{i+1,j+1} & I_{i,j+1} & I_{i-1,j+1} & \cdots \\ \cdots & I_{i+1,j} & I_{i,j} & I_{i-1,j} & \cdots \\ \cdots & I_{i+1,j-1} & I_{i,j-1} & I_{i-1,j-1} & \cdots \end{bmatrix}.$$

(12)

The 4-step (2x2) SC-PSA kernel is given equivalently as

$$\left[(-1)^{(i+j)}\varphi_{i,j}+pm_{i,j}\right]\approx\tan^{-1}\left[\frac{I_{i-1,j}-I_{i,j-1}}{-I_{i,j}+I_{i-1,j-1}}\right]=\arg\left\{\begin{bmatrix}-1 & i \\ -i & 1\end{bmatrix}*I_{i,j}\right\}.$$

(13)

Thus, one must remove the pixelated-carrier ($pm_{i,j}$), and correct the phase-signs $[(-1)^{i+j}\varphi_{i,j}]$. The resulting phase is,

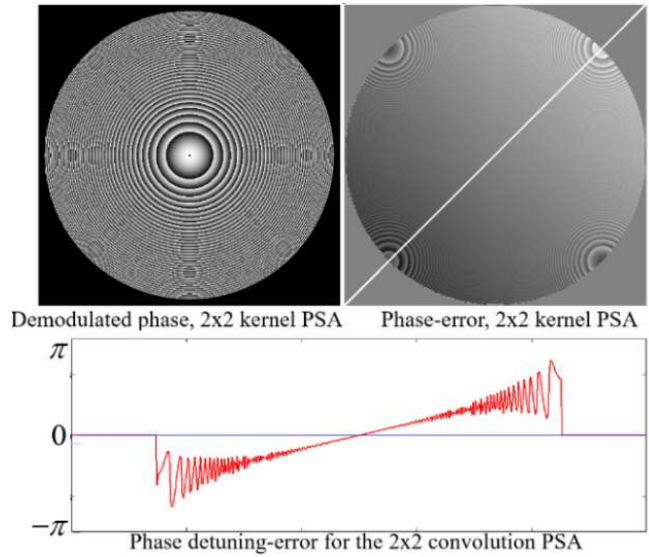

Fig. 9. Spatial convolution demodulation by the smooth-edge 2x2 kernel. We show the wrapped phase, and the phase-error image and along the white crossing line (red graph). The tilt is introduced by the 2x2 kernel itself, it is not a part of the measuring phase.

The sign changes and pixelated carrier in Eq. (13) does not appear in the 4DTC papers [9-15]. The 2x2 demodulation is an approximation to the error-free phase of wideband fringes, that is why the sign ($\approx$) is used.

### 14.2 Erroneous demodulation by 3x3 spatial phase-shifting complex convolution kernel

Kimbrough state that "*the circular-9 formula is*" (Eq. (35) in [13]),

$$\left[(-1)^{(i+j)}\varphi_{i,j} + pm_{i,j}\right] \approx \tan^{-1}\left[\frac{2I_{i,j-1} + 2I_{i,j+1} - 2I_{i+1,j} - 2I_{i-1,j}}{-I_{i-1,j-1} - I_{i+1,j+1} + 4I_{i,j} - I_{i+1,j-1} - I_{i-1,j+1}}\right]. \tag{14}$$

Equivalently, this may be expressed as a 3x3 complex-valued quadrature convolution filter,

$$\left[(-1)^{(i+j)}\varphi_{i,j} + pm_{i,j}\right] \approx \arg\left\{\begin{bmatrix} -1 & -2i & -1 \\ 2i & 4 & 2i \\ -1 & -2i & -1 \end{bmatrix} * I_{i,j}\right\}. \tag{15}$$

Thus, one must remove the pixelated-carrier and the phase-sign changes; this was not reported in [9-15].

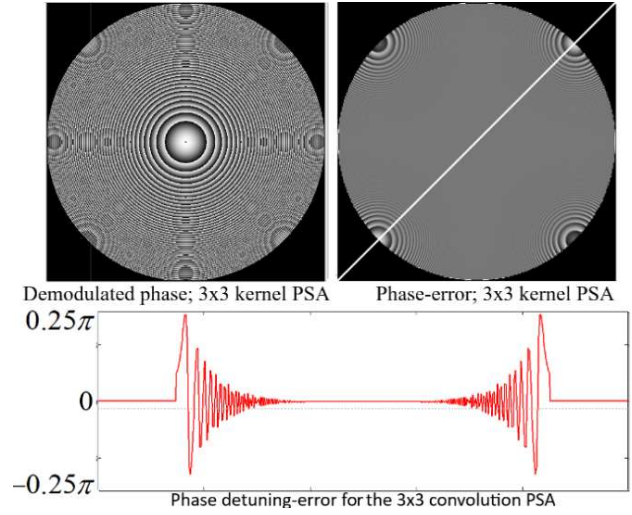

Fig. 10. We show the phase demodulation and phase-error images The kernel used is the smooth-edge 3x3 SC-PSA. Also, we show the crossing line to visualize the phase detuning-error (in red); the white-line crosses the gray-image phase-error.

The phase-error arise because the complex-conjugate signal is not fully filtered-out by the smooth-edge cutoff 3x3 kernel SC-PSA [9-15]; hence the use of the approximation sign $(\approx)$.

## 14. Detuning due to crosstalking among conjugate spectra of the 4DTC's convolution algorithms

To the best of our knowledge, this is an original contribution of this work [9–15]. Throughout 2004–2009, 4DTC demodulation methods remained strictly in the spatial domain. It appears that 4DTC was unaware that their 2×2 and 3×3 SC-PSAs were, in fact, high-pass complex quadrature filters—an insight that was never published by 4DTC [9–15]. Here, we provide the missing spectral analysis and explicitly identify the high-pass complex kernels, which were foundational to their early demodulation algorithms [9-15].

### 15.1 The FTF of the 2x2 and 3x3 convolution algorithms used by 4TDC within 2004-2009 [9-15]

We have seen that 4DTC's 2x2 and 3x3 complex high-pass SC-PSAs kernels are the following,

$$\text{4-step PSA;} \quad hp_{2x2} = \begin{bmatrix} -i & -1 \\ 1 & i \end{bmatrix}; \qquad \text{circular 9-step PSA;} \quad hp_{3x3} = \begin{bmatrix} -1 & -2i & -1 \\ 2i & 4 & 2i \\ -1 & -2i & -1 \end{bmatrix}. \tag{16}$$

Their Fourier transform is,

$$HP_{2x2}(u,v) = \mathcal{F}\{hp_{2x2}\} = -i - e^{-iu} + e^{-iv} + ie^{-i(u+v)} \quad ; \tag{17a}$$

$$HP_{3x3}(u,v) = \mathcal{F}\{hp_{3x3}\} = 4 - 2[\cos(u+v) + \cos(u-v)] + 4i[\cos(u) - \cos(v)]. \tag{17b}$$

The 4DTC's 2x2 and 3x3 SC-PSAs are,

$$\left[(-1)^{i+j}\varphi_{i,j} + pm_{i,j}\right] \approx \arg[hp_{2x2} * I(x,y)]; \tag{18a}$$

$$\left[(-1)^{i+j}\varphi_{i,j} + pm_{i,j}\right] \approx \arg[hp_{3x3} * I(x,y)]. \tag{18b}$$

The FTFs are $HP_{2x2}(u,v)$ and $HP_{3x3}(u,v)$ as the next two figures show. Due to the 2x2 and 3x3 smooth-edge filters, the phase-error is due to cross-tacking among the spectral components of $I(u,v) = \mathcal{F}\{I(x,y)\}$.

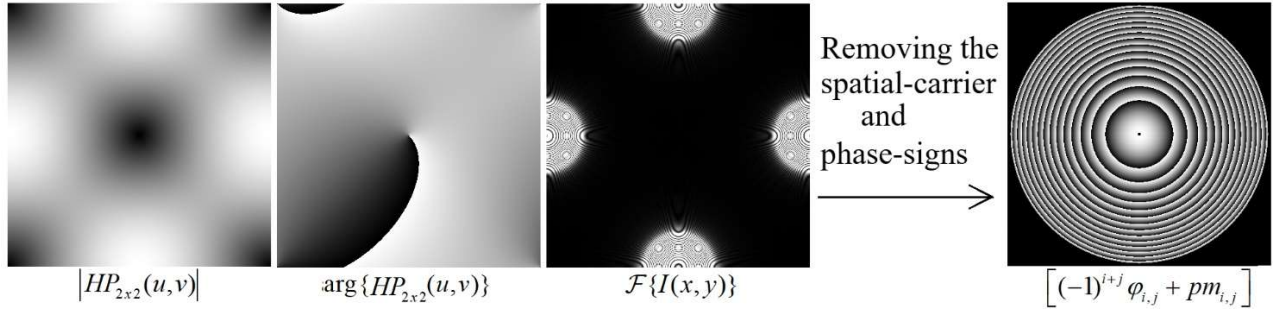

Fig. 11. Smooth-edge cutoff 2x2 kernel producing crosstalking of conjugate spectra, and its phase [9-15]. This phase may have small phase detuning error, given that the pixelated fringe spectrum has narrow bandwidth.

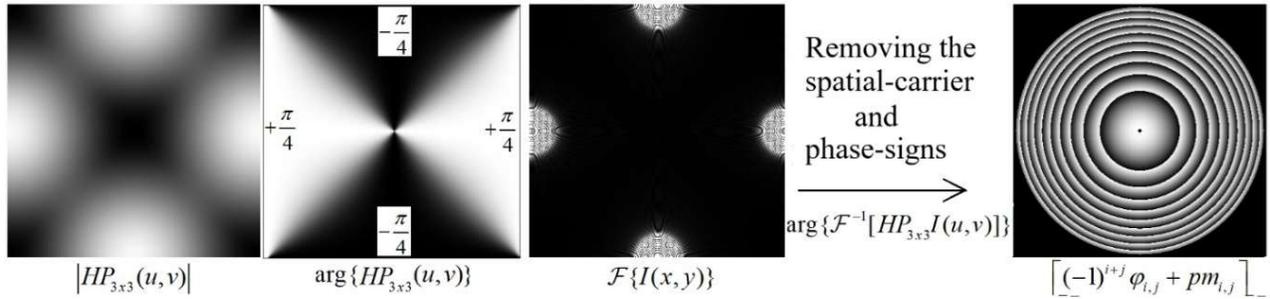

Fig. 12. Smooth-edge cutoff 3x3 kernel allowing crosstalking of conjugate spectra giving erroneous phase demodulation [9-15]. In this particular case no phase error is present due to the small bandwidth of the pixelated fringe spectrum.

The above two figures, demonstrate that the 2×2 and 3×3 smooth-edge quadrature filters introduce cross-talk between the conjugate spectral components—signals that must remain spectrally separated to enable error-free demodulation. This spectral overlap leads to unacceptable erroneous phase error when applied to wideband pixelated interferograms [13], [9–15]. Importantly, 4DTC never published any spectral analysis of their 2×2 and 3×3 SC-PSAs during 2004-2009 [9–15]. In fact, their work gives no indication that they understood the frequency response of their own convolution algorithms; they never reported their corresponding FTFs [9-15], as we have done in this section. This lack of spectral insight suggests that 4DTC had no clear understanding of the internal spectral behavior of their 2×2 and 3×3 CS-PSAs. For this reason, the analysis presented here is critical—it serves as compelling circumstantial evidence supporting the claim that 4DTC's 2010 work may have appropriated key ideas from Servin's research still in peer-review [17].

## 15. Servin's non-crosstalking spectra demodulation of wideband pixelated interferograms

For the reader's convenience, we resume Servin's paper [17]. The pixelated interferogram model is,

$$I(x,y) = a(x,y) + b(x,y)\cos[\varphi(x,y) + pm(x,y)]. \tag{19}$$

To create these fringes, the reference wavefront must be [17].

$$R(x,y) = e^{-i\,pm(x,y)}. \tag{20}$$

The crucial Ansatz was to consider that the fringe-carrier product may spectrally separate the desired signal.

$$I(x,y)R(x,y) = \{a + b\cos[\varphi + pm]\}e^{-i\,pm}. \tag{21}$$

A second Ansatz is that a step-edge lowpass filter $LP_{circ}(u,v)$ will separate the desired spectrum from its conjugate,

$$\frac{b}{2}e^{i\varphi} = \mathcal{F}^{-1}\{LP_{circ}\mathcal{F}[I e^{-i\,pm}]\}; \tag{22}$$

The Ansatz were confirmed by demodulating the pixelated interferogram as the next figure shows.

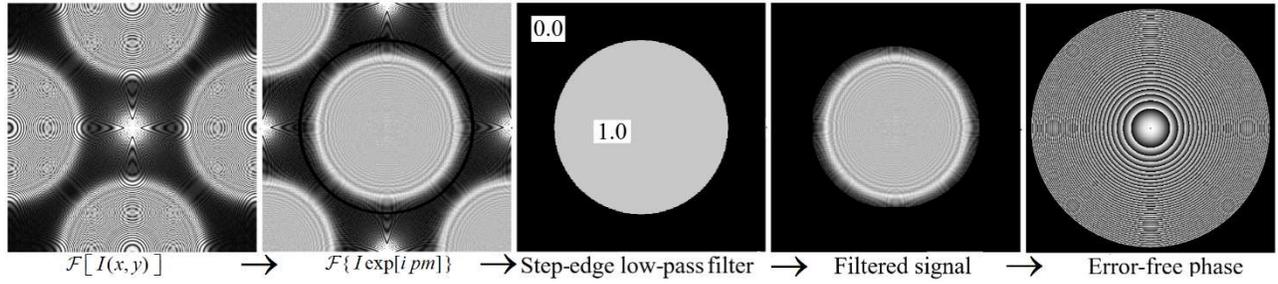

Fig. 13. The complex-valued fringe-carrier product Ansatz. The desired signal is separated by a step-edge low-pass filter [17].

The energy outside the low frequency signal must be deleted by the step-edge filter $LP_{circ}(u,v)$. This was stated by Servin et al.´s paper [17] as follow: "*Note that, phase detuning error arises because some undesired signal from the conjugate spectrum leaks into the desired analytical signal [9]. In our method this is not possible because we have filtered-out completely this conjugate spectrum in the Fourier domain.*" (Italics mine) [17].

## 16. Unexplained shift from 2x2 and 3x3 high-pass [9-15] to 2x2 and 3x3 low-pass kernel filters [18]

To our knowledge, the material in this section has not been published, thus it represents a new contribution of this work.

It appears that 4DTC did not recognize that their 2×2 and 3×3 SC-PSAs [9–15] were actually quadrature filters—otherwise, they would have reported their Fourier Transfer Functions (FTFs), which they never did [9-15]. This sudden change from complex-valued high-pass kernels, to real low-pass kernels are,

$$hp_{2x2} = \overbrace{\begin{bmatrix} -i & -1 \\ 1 & i \end{bmatrix}}^{Highpass\ filter} \to lp_{2x2} = \overbrace{\begin{bmatrix} 1 & 1 \\ 1 & 1 \end{bmatrix}}^{Lowpass\ filter}; \quad hp_{3x3} = \overbrace{\begin{bmatrix} -1 & -2i & -1 \\ 2i & 4 & 2i \\ -1 & -2i & -1 \end{bmatrix}}^{Highpass\ filter} \to lp_{3x3} = \overbrace{\begin{bmatrix} 1 & 2 & 1 \\ 2 & 4 & 2 \\ 1 & 2 & 1 \end{bmatrix}}^{Lowpass\ filter}. \quad (23)$$

$$\underbrace{\phantom{xxxxx}}_{2004-2009\ [9-15]} \underbrace{\phantom{xxxxx}}_{2010\ [18]} \qquad \underbrace{\phantom{xxxxx}}_{2004-2009\ [9-15]} \underbrace{\phantom{xxxxx}}_{2010\ [13]}$$

As reminder, their Frequency Transfer Functions (FTFs) are,

$$\begin{aligned}
HP_{2x2} &= \mathcal{F}\{hp_{2x2}\} = -i - e^{-iu} + e^{-iv} + ie^{-i(u+v)}; \\
HP_{3x3} &= \mathcal{F}\{hp_{3x3}\} = 4 - 2[\cos(u+v) + \cos(u-v)] + 4i[\cos(u) - \cos(v)]; \\
LP_{2x2} &= \mathcal{F}\{lp_{2x2}\} = 4e^{0.5i(u+v)}\cos(0.5u)\cos(0.5v); \\
LP_{3x3} &= \mathcal{F}\{lp_{3x3}\} = [\cos(0.5u)\cos(0.5v)]^2.
\end{aligned} \quad (24)$$

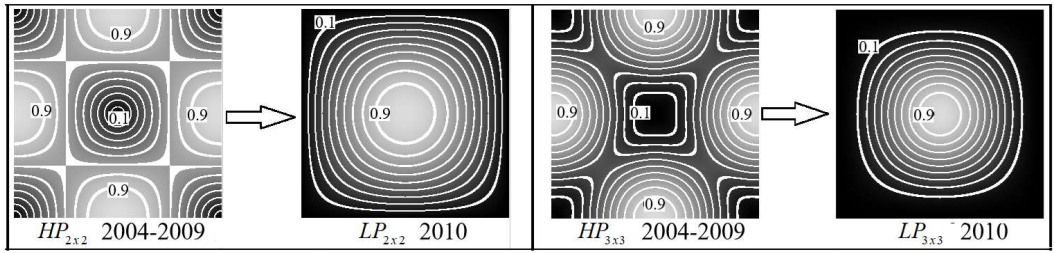

Fig. 14. Spectral modulus for the 2x2 and 3x3 high-pass filtering [9-15] and 2x2 and 3x3 low-pass filtering demodulation [18].

This figure displays contour plots comparing the modulus of 4DTC's published phase-demodulation filters:
- High-pass SC-PSA filters $HP_{2x2}$ and $HP_{3x3}$ used from 2004 to 2009 [9–15], and
- Low-pas*s* Fourier-filters $LP_{2x2}$ and $LP_{3x3}$ abruptly introduced in 2010 [18].

The abrupt change from high-pass SC-PSAs (employed for six years) to Fourier-domain low-pass demodulation filters strongly suggest that 4DTC's 2010 approach [18] was not an independent development, but likely derived from Servin's unpublished work [17].

## 17. 4DTC's fringe-carrier product plus 3x3 low-pass filtering Fourier-demodulation [17,18]

For exposition clarity, we use a similar notation to K&M [18]. Their Eq. (1), the interferogram formula is [18],

$$I_{det}(x,y) = I_{avg}(x,y) + vI_{avg}(x,y)\text{Cos}[\theta_w(x,y) + \phi_c(x,y)]; \quad a = I_{avg}; b = vI_{avg}. \quad (25)$$

The measuring phase is $\theta_w$ and the pixelated carrier is $\phi_c$. The unusual notation $\text{Cos}[\cdot]$ instead of $\cos[\cdot]$ is in [18].

Their K&M's fringe-carrier complex product is [18],

$$\tilde{I}_{HET} \equiv I_{det} e^{-i\phi_c}. \tag{26}$$

Note the sudden change to complex analysis, never used before [9-15]. The K&M low-pass Fourier-filtering is,

$$\tilde{I}_{alg} = \mathcal{F}^{-1}\left\{H_f \mathcal{F}\left[I_{det} e^{-i\phi_c}\right]\right\}. \tag{27}$$

For example, using the $H_f(u,v) = LP_{3x3}(u,v)$ gives erroneous phase-demodulation for wideband fringes [18],

$$\varphi_{i,j} \approx \arg\left[\mathcal{F}^{-1}\left\{LP_{3x3}\mathcal{F}\left[I_{det}e^{-i\phi_c}\right]\right\}\right]. \tag{28}$$

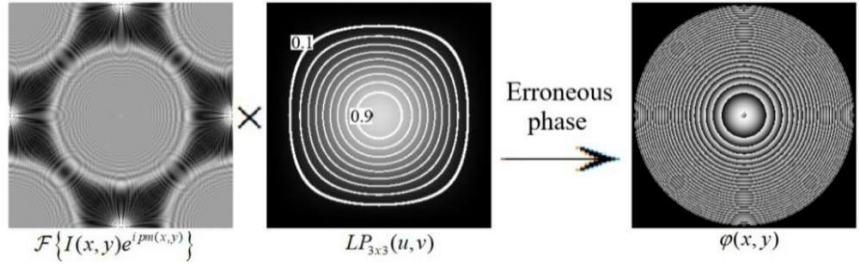

Fig. 15. Cross-talking between wideband conjugate spectra filtered by the 3x3 kernel with smooth-edge cutoff used in [18]. As can be seen this cross-talking can be repaired simply changing this filter to a step-edge filter [17] as the next figure shows.

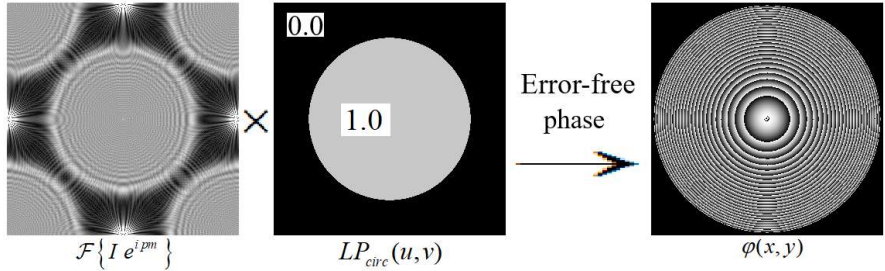

Fig. 16. Servin et al.'s symmetrical bandwidth step-edge low-pass Fourier-filter used in [17]. It is trivial to use this step-edge filter instead of the 3x3 smooth-edge low-pass kernel in the preceding figure and obtain error-free phase demodulation.

Why did 4DTC stop short of using the straightforward step-edge filtering employed by Servin et al. [17] to improve their flawed 3x3 SC-PSA [18]? We believe Kimbrough and Miller (K&M) intentionally avoided this filter to make their SPIE-7790 paper [18] appear independent from Servin's [17]. Had they used both: The same fringe-carrier product, and the same step-edge filter for error-free demodulation, the resemblance would have been too evident—clearly exposing their reliance on Servin's methodology. Moreover, in 2010, K&M still promoted their flawed 2×2 and 3×3 filters (now as Fourier low-pass filters), likely in an effort to remain aligned with the abstract they had submitted to the SPIE-7790 conference by April 10 [18].

## 18. Phase-error due to crosstalking of conjugate spectra by the 2x2 and 3x3 kernels [13,18]

The Fourier-based analysis presented in this section was not previously disclosed by 4DTC [9–15,18,19], and thus constitutes a new contribution of this work. Critically, the amount of phase detuning-error analyzed here remains identical for both 2x2 and 3x3 high-pass filters reported in [9–15] and their later 2×2 and 3×3 low-pass Fourier filters introduced in 2010 [18]—this major methodological shift (from low-pass to Fourier high pass filtering) was implemented without any acknowledgment, explanation, analysis, or discussion of its implications in 4DTC's [18].

Let us start by consider a tilted measuring wavefront and its corresponding pixelated interferogram as,

$$E_{obj} = |E_{obj}|e^{i\varphi} = |E_{obj}|e^{i\sigma_0(i+j)}; \qquad \sigma_0 \in [0, 0.5\pi);$$

$$I(pm) = a + b\cos[\varphi + pm] = a + \frac{b}{2}e^{i[\varphi+pm]} + \frac{b}{2}e^{-i[\varphi+pm]}. \tag{29}$$

For ease of analysis, we use Servin's fringe-carrier product as [17],

$$\mathcal{F}\{I(pm)e^{ipm}\} = \mathcal{F}\{ae^{ipm}\} + \frac{b}{2}\delta_1(u-\sigma_0, v-\sigma_0)e^{i\varphi} + \frac{b}{2}\delta_2[u-\pi+\sigma_0, v-\pi+\sigma_0]e^{-i\varphi}. \tag{30}$$

Spectral crosstalking between the desired signal $\delta_1(u-\sigma_0, v-\sigma_0)e^{i\varphi}$ and its conjugate $\delta_2[u-\pi+\sigma_0, v-\pi+\sigma_0]e^{-i\varphi}$.

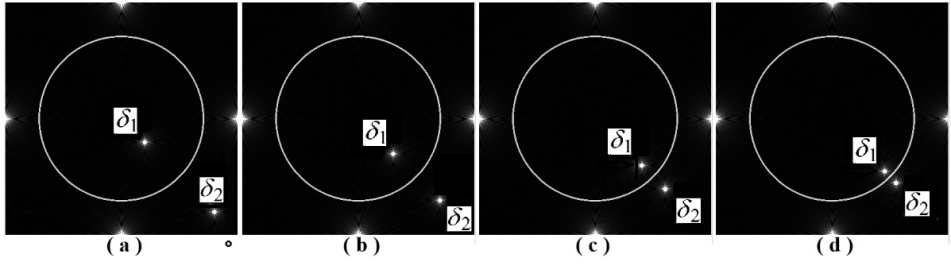

Fig. 17. The circled bandwidth is, $u^2+v^2=(\pi/\sqrt{2})^2$. The desired delta is $\delta_1$, and its undesired wrapped conjugate is $\delta_2$

The fringe-carrier product spectra, while $\sigma_0 \in \{0.2\pi, 0.3\pi, 0.4\pi, 0.45\pi\}$. The erroneous demodulated phase is,

$$\varphi \approx \arg\{LP_{2\times 2}(\sigma_0,\sigma_0)e^{i\varphi} + LP_{2\times 2}[\pi-\sigma_0,\pi-\sigma_0]e^{-i\varphi}\};$$
$$\varphi \approx \arg\{LP_{3\times 3}(\sigma_0,\sigma_0)e^{i\varphi} + LP_{3\times 3}[\pi-\sigma_0,\pi-\sigma_0]e^{-i\varphi}\}; \quad \sigma_0 \in [0, 0.5\pi). \tag{31}$$

A diagonal-cut of $LP_{2\times 2}(u,u)$ and $LP_{3\times 3}(u,u)$, and the testing fringe deltas 1, and 2 are.

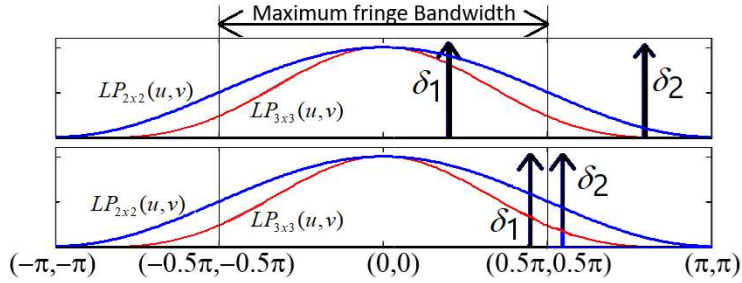

Fig. 18. Spectral crossing-line between $\delta_1$ and $\delta_2$ (Fig. 17) for the 2x2 and 3x3 low-pass PSA filters [9-15,18].

The detuning amplitude error may be defined as [20],

$$D(\sigma_0) = \frac{Conjugate\ Signal}{Desired\ Signal} = \frac{|LP_{N\times N}[\pi-\sigma_0,\pi-\sigma_0]|}{|LP_{N\times N}(\sigma_0,\sigma_0)|}; N \in \{2,3\}. \tag{32}$$

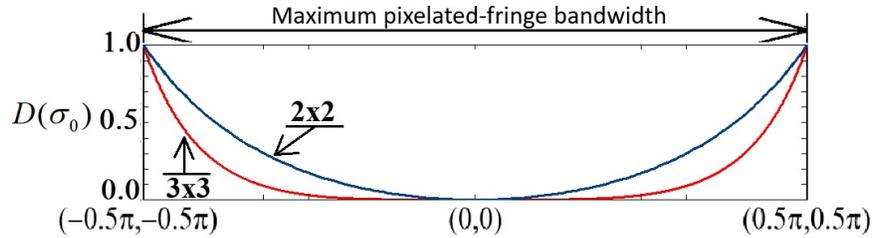

Fig. 19. Detuning amplitude error $D(\sigma_0)$ due to crosstalking for the 4DTC's 2x2 (in red), and 3x3 (in blue) algorithms [9-15].

One can see that of $LP_{2\times 2}(u,u)$ and $LP_{3\times 3}(u,u)$, can only demodulate with low phase-error, low bandwidth fringes.

The following figure shows the $LP_{2\times 2}(u,u)$ and $LP_{3\times 3}(u,u)$ [18] compared with the step-edge filter used in [17].

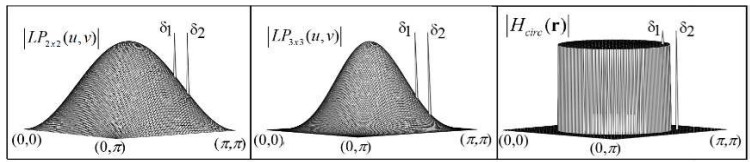

Fig. 20. Spectral crosstalking between the desired $\delta_1$ and its conjugate $\delta_2$ for the 4DTC's 2x2, 3x3 algorithms (in the Fourier-domain) and the non-crosstalking step-edge filter. Delta $\delta_1$ is the desired signal and its undesired conjugate is $\delta_2$.

Why did Kimbrough and Miller continue to rely on their flawed 2×2 and 3×3 kernels in [18]? They wrote: "*If a symmetric filter with a step edge cutoff were possible, the dashed line in each spectrum indicates the maximum spatial frequency to prevent aliasing for a wavefront with only tilt and a flat background intensity*" (italics mine) [18]. Yet, implementing a step-edge Fourier filter is technically trivial.

Their strategic omission of the straightforward step-edge low-pass filtering conveniently aligned with the abstract submitted to the SPIE-7790 conference prior to April 10, 2010 [18], which made no reference to the fringe-carrier product or to Fourier-filtering demodulation. This calculated exclusion—despite its technical simplicity—strongly indicates that Kimbrough and Miller had the conscious intention to frame their paper [18] as an independent development from Servin's [17]. Had they incorporated both the fringe-carrier product and step-edge Fourier-filtering—essential components of Servin's error-free solution—their dependence on Servin's still-unpublished manuscript [17] would have been difficult to deny. We must also be aware that, the Fourier-demodulation method presented in [18] continued using smooth-edged 2×2 and 3×3 kernel filters, thereby reproducing the very same phase-errors reported in their earlier works [13].

## 19. Phase demodulation using K&M's own simplified pixelated-fringe model carriers

In [18] K&M propose a new and equivalent fringe model using two carriers $\text{Cos}[\pi x]$, $\text{Cos}[\pi y]$ as,

$$I_{KM} = a + b\,\text{Cos}\!\left[\theta_w + \phi_c + \frac{\pi}{4}\right] = a + \frac{\sqrt{2}}{2}b\{\text{Cos}[\theta_w]\text{Cos}[\pi x] + \text{Sin}[\theta_w]\text{Cos}[\pi y]\}. \tag{33}$$

This new fringe model was highlighted by K&M as: "*The simplified result is given in equation 5.*" [18]. Why K&M did not use their simplified fringe's carriers for demodulation? This would give them the following demodulation formula,

$$|Z|e^{i(\theta_w+\frac{\pi}{4})} \approx \mathcal{F}^{-1}\!\left[LP_{3x3}\mathcal{F}\{I_{KM}[\text{Cos}(\pi x)+i\,\text{Cos}(\pi y)]\}\right]; \quad i=\sqrt{-1}. \tag{34}$$

This demodulator *does not appear in* [18], *it is a contribution of this work*. To my surprise instead of using their own carriers $[\text{Cos}(\pi x),\text{Cos}(\pi y)]$ for demodulation, K&M uses the CIO's fringe-carrier product $I_{KM}\,e^{i\phi_c}$ approach [17],

$$|Z|e^{i\theta_w} \approx \mathcal{F}^{-1}\!\left[LP_{3x3}\mathcal{F}\{I_{KM}\,e^{i\,pm}\}\right]. \tag{35}$$

Both demodulators yield exactly the same phase result $\theta_w(x,y)$. Had K&M actually employed their originally proposed carriers $[\text{Cos}(\pi x),\text{Cos}(\pi y)]$ for demodulation, it might have partially obscured the clear evidence that 4DTC's demodulation method was directly derived from [17]. But given that this new phase demodulator still uses the 3×3 low-pass kernel [18] it exhibits the very same demodulation errors as their earlier high-pass kernels [9-15], [13].

## 20. Decoupling the crosstalking between conjugate spectra by modifying the 3x3 SC-PSA

As far as we know, the material in this section was not published before, so it represents a contribution of this work.

**Case-1.** From their own 3x3 algorithm, 4DTC could have improved their 3x3 quadrature-demodulator as.

$$HP1(u,v) = \{HP_{3x3}(u,v),\quad if\,|HP_{3x3}|\geq 4\,;\quad 0,\quad otherwise\};$$
$$\left[(-1)^{i+j}\varphi_{i,j} - pm_{i,j}\right] \approx \arg\!\left[\mathcal{F}^{-1}\{HP1(u,v)\,I(u,v)\}\right]; \quad I(u,v)=\mathcal{F}\{I(x,y)\}. \tag{36}$$

The $HP1$ filter eliminates the cross-talking of the spectral components, still remaining low filter's phase distortion.

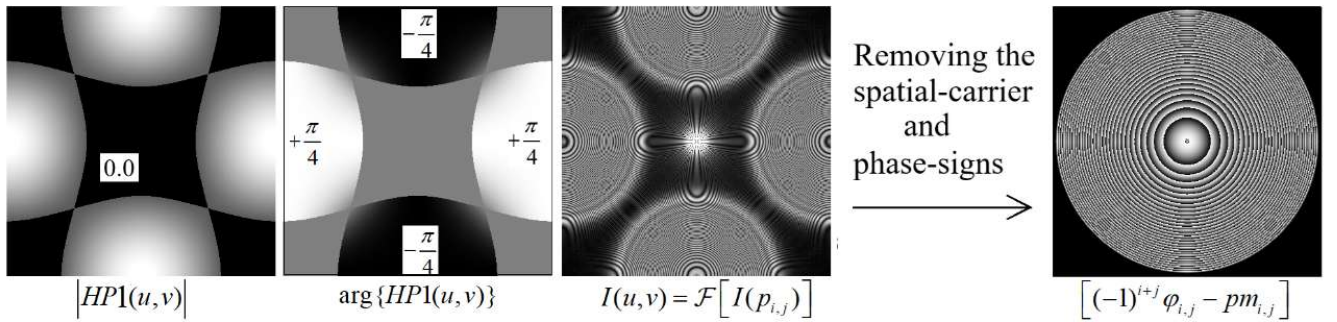

Fig. 21. Non-crosstalking (isolated) conjugate signals achieved using a step-edge Fourier filter derived from the smooth-edge cutoff of the 3×3 SC-PSA via a thresholding operation. The spectral energy which connects the conjugate lobes have been forced to zero, effectively decoupling them. However, the phase response of this filter remains non-flat, exhibiting an undesired smooth roll-off near the cutoff, which introduces small non-zero phase distortion.

**Case-2.** One may still improve this to the widest bandwidth filter's magnitude, still having small phase distortion.

$$HP2(u,v) = |HP_{circ}(u,v)| HP_{3x3}(u,v);$$
$$\left[(-1)^{i+j}\varphi_{i,j} - pm_{i,j}\right] \approx \arg\left[\mathcal{F}^{-1}\{HP2(u,v)I(u,v)\}\right].$$
(37)

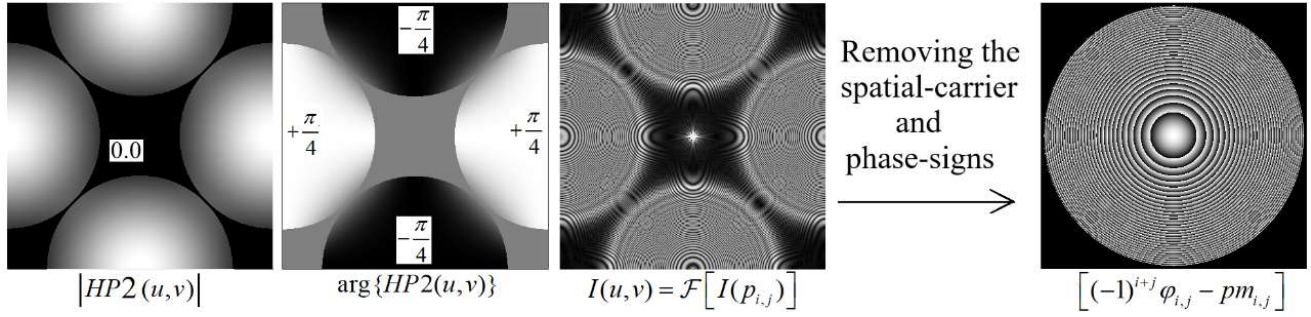

Fig. 22. Widest non-crosstalking quasi-step-edge quadrature filter obtained from the 4DTC's 3x3 algorithm [9-15]. Still the filter's phase-response is not flat, so it still has undesired smooth-phase roll-off cutoff response.

**Case-3.** Finally, the step-edge error-free quadrature solution, similar to that of Servin's [17] is the following,

$$HP_{circ}(u,v) = e^{i\frac{\pi}{4}}[LP_{circ}(u-\pi,v) + LP_{circ}(u+\pi,v)] + e^{-i\frac{\pi}{4}}[LP_{circ}(u,v-\pi) + LP_{circ}(u,v+\pi)];$$
$$\left[(-1)^{i+j}\varphi_{i,j} - pm_{i,j}\right] = \arg\left[\mathcal{F}^{-1}\{HP_{circ}(u,v)I(u,v)\}\right]; \quad \text{Error-free wideband phase-demodulation.}$$
(38)

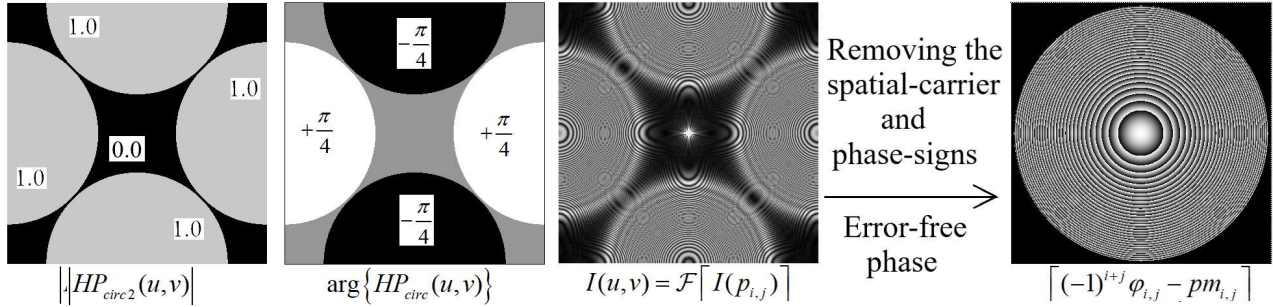

Fig. 23. Widest non-cross-talking step-edge quadrature filter obtained from the 4DTC's 3x3 algorithm [9-15]. This Fourier-filter has now flat step-edge magnitude and phase responses, and it is now error-free and fully equivalent to that of Servin et al. [17].

These progressively improved quadrature demodulators—Cases 1 through 3—constitute a natural evolution from the 3×3 smooth-edge spatial-convolution PSAs (SC-PSAs) developed by 4DTC between 2004 and 2009 [9–15]. In our own analysis, we followed this logical trajectory to reach an error-free demodulation framework directly from the foundations laid by 4DTC's earlier work [9–15]. Had 4DTC properly analyzed the frequency response of their own 2×2 and 3×3 SC-PSA filters, they could have derived a sequence of better high-pass quadrature filters that incrementally reduced phase errors in wideband pixelated interferometry. Such a development—entirely consistent with their prior methodology—would have offered a clear and independent route to improved performance, eliminating any grounds for questioning the originality of their SPIE-7790 paper [18].

However, they failed to take this progressive path, likely because the 4DTC lacked understanding of the FTFs associated with their spatial-convolution algorithms. As mentioned, knowledge of these FTFs would have naturally guided them to refine their 3x3 smooth-cutoff quadrature filter, towards the step-edge error-free quadrature-filter $HP_{circ}(u,v)$.

As analyzed in this section, 4DTC bypassed their own prior trajectory [9–15] and abruptly adopted the core concepts introduced by Servin et al. [17]—namely, the fringe-carrier product followed by low-pass Fourier filtering. This methodological shift occurred just weeks after Servin's manuscript was submitted [17], following six years of stagnation during which 4DTC had dismissed Fourier-based processing [9–15]. The combination of timing, conceptual overlap, methodological mirroring, and their previous disregard for spectral analysis raises serious and well-founded concerns about the independence and originality of their 2010 SPIE contribution [18].

## 21. Ideas concepts and methods likely plagiarized from Servin's work [17] still in peer-review

As previously discussed, plagiarism extends beyond the reuse of exact wording to include the unacknowledged appropriation of paradigm-shifting ideas, methods, or lines of reasoning. In this context, we now present a brief summary of the key concepts introduced by Servin's manuscript, "*Error-Free Demodulation of Pixelated Carrier Frequency Interferograms*" [17] that were highly probable copied by K&M in their SPIE 7790 paper [18]. These include:

1) Mathematical model for pixelated spatial-carrier interferograms,

$$I(x,y) = a(x,y) + b(x,y)\cos[\varphi(x,y) + pm(x,y)]. \quad \text{(Servin's [17])};$$

$$I_{det}(x,y) = a(x,y) + b(x,y)\text{Cos}[\theta_w(x,y) + \phi_c(x,y)]; \quad \text{(Kimbrough's [18])}.$$

After presumably gaining access to Servin's manuscript [17] during peer-review, 4DTC appeared to realize that advancing their previously six-years stagnant theory of pixelated interferometry—from 2004 to 2009—required a formal mathematical model of pixelated fringes. However, no such fringe modeling had been published by 4DTC prior to 2010 [9–15], nor was in any reference in the sources cited in [18].

2) Pixelated interferogram Fourier spectrum.

$$I(u,v) = \mathcal{F}\{I(x,y)\} = \mathcal{F}\{a(x,y) + b(x,y)\cos[\varphi(x,y) + pm(x,y)]\}. \quad \text{(Servin's [17])};$$

$$I_{det}(u,v) = \mathcal{F}\{I_{det}(u,v)\} = \mathcal{F}\{a(x,y) + b(x,y)\text{Cos}[\theta_w(x,y) + \phi_c(x,y)]\}; \quad \text{(Kimbrough's [18])}.$$

Before 2010, 4DTC had never published a Fourier spectrum of a pixelated interferogram [9–15]. However, just seven weeks after the submission of Servin's manuscript [17], 4DTC unexpectedly introduced spectral analysis of pixelated fringes in their SPIE-7790 presentation [18]. From the spectrum alone, it is not readily apparent how to separate the two conjugate components. None of 4DTC's earlier publications [9–15], nor any references cited in [18], included spectral analysis of pixelated fringes. Notably, terms such as "Fourier analysis" or "Fourier-domain phase demodulation" are entirely absent from the abstract of their SPIE-7790 contribution [18].

3) Synchronous-cophased fringe-carrier product as Servin did in [17],

$$I \times R = \{a + b\cos[\varphi + pm]\}e^{-ipm}. \quad \text{(CIO's [17])};$$

$$\tilde{I}_{HET} \equiv I_{det} e^{-i\phi_c}; \quad \text{(4DTC's [18])}.$$

After likely obtaining access to [17] during the peer-review process, 4DTC recognized the utility of the fringe-carrier cophased product—an approach not documented in any of their publications prior to 2010 [9–15]. In their paper [18], they referred to this quantity as the "heterodyne product," a misnomer, as previously discussed. Furthermore, the term "heterodyne product" does not appear in the abstract of their SPIE-7790 paper [18].

4) Fourier spectrum of the synchronous-cophased fringe-carrier product as written in [17],

$$\mathcal{F}\{I \times R\} = \mathcal{F}\{[a + b\cos(\varphi + pm)]e^{-ipm}\}. \quad \text{(CIO's [17])};$$

$$\mathcal{F}\{\tilde{I}_{HET}\} = \mathcal{F}\{I_{det} e^{-i\phi_c}\}; \quad \text{(4DTC's [18])}.$$

Just seven weeks after Servin's manuscript submission [17], 4DTC reported a finding that identified the fringe-carrier cophased product spectrum as the critical element for phase demodulation of pixelated fringes [18]. This cophased product effectively separates the fringe spectrum into two distinct components: the desired baseband signal and its high-frequency conjugate. There is no indication that 4DTC had previously recognized or explored this approach, as their publications from 2004 to 2009 [9–15] contain no reference to or discussion of such a fringe-carrier cophased product.

5) Low-pass Fourier-filtering of the fringe-carrier product strategy

$$\frac{b}{2}e^{i\varphi} = \mathcal{F}^{-1}\{LP_{circ}\mathcal{F}[Ie^{-ipm}]\}; \quad \text{CIO's pixelated demodulator [17]};$$

$$\tilde{I}_{alg} = \mathcal{F}^{-1}\{LP_{3x3}\mathcal{F}[I_{det} e^{-i\phi_c}]\}; \quad \text{4DTC's pixelated demodulator [18]}.$$

Just weeks after Servin's submission [17], 4DTC abruptly adopted these key concepts—(1) pixelated fringe modeling, (2) spectral analysis of the fringes, (3) cophased fringe-carrier product, (4) spectral cophased product analysis, and (5) Fourier low-pass filtering cophased demodulation—all of which were absent from their prior works [9–15], yet became central to their 2010 approach [18], suggesting reliance on [17] rather than independent development. Notably, their 2×2 and 3×3 low-pass kernels in [18] retained the same phase errors as their earlier 2×2 and 3×3 high-pass kernels [9–15], a critical oversight that was neither acknowledged, addressed, nor analyzed in their SPIE-7790 work [18], leaving post-2010 readers blind to this paradigm shift and without much-needed clarification.

The only concept 4DTC did not appropriate from Servin's manuscript [17] was the step-edge low-pass Fourier filter. In 2010, 4DTC persisted with their flawed 2×2 and 3×3 SC-PSA kernels [9–15,18], despite knowing by then that these filters were intrinsically defective. As previously explained, their smooth-edge cutoff mixes conjugate components in the fringe-carrier spectrum, causing phase errors. Error-free demodulation requires complete separation of these spectral terms [17].

Yet 4DTC could not incorporate Servin's step-edge filter without contradicting their SPIE abstract submitted before April 2010 (timeline Fig. 3; abstract Fig. 4), which still promoted 2×2 and 3×3 filters. Adopting the new method would have exposed a sudden and even more difficult-to-explain shift.

Rather than acknowledge Servin's practical solution, K&M wrote in [18]: "*If a symmetric filter with a step edge cutoff were possible, the dashed line in each spectrum indicates the maximum spatial frequency to prevent aliasing for a wavefront with only tilt and a flat background intensity.*" (italics mine) This misleadingly implies that step-edge filters are unfeasible. In fact, these filters are standard in digital signal processing (see Servin's, Digital Interferometry [20]).

## 22. Various different alternatives to phase-demodulate pixelated interferograms

Just as remainder, the 2x2 and 3x3 spatial-convolution phase-shifting algorithms (SC-PSAs) $(\cdot * \cdot)$ are [9-15],

$$(a); \quad \left[(-1)^{i+j}\varphi_{i,j} - pm_{i,j}\right] \approx \arg\left[hp_{2x2} * I(x,y)\right];$$
$$(b); \quad \left[(-1)^{i+j}\varphi_{i,j} - pm_{i,j}\right] \approx \arg\left[hp_{3x3} * I(x,y)\right] \tag{39}$$

Being $(\cdot * \cdot)$ the convolution operation. From 2004 to 2009, 4DTC used just these two SC-PSAs demodulators [9–15].

By 2010, 4DTC had several clear and technically feasible paths to achieve more accurate phase demodulation by simply improving upon their own 3×3 SC-PSA algorithm used between 2004 and 2009 [9–15]. Each of these alternatives emerged as a natural and independent improvements within their previous quadrature-filtering methodology. This would have required nothing more than the spectral analysis of their 3×3 convolution kernel—yet they failed to pursue this route. Now let us construct these demodulators with progressively reduced phase error as,

$$(c); \quad \left[(-1)^{i+j}\varphi_{i,j} - pm_{i,j}\right] \approx \arg\left[\mathcal{F}^{-1}\{HP1\, I(u,v)\}\right]; \quad HP1 = \begin{cases} HP_{3x3}, & \text{if } |HP_{3x3}| \geq 4 \\ 0, & \text{otherwise} \end{cases};$$
$$(d); \quad \left[(-1)^{i+j}\varphi_{i,j} - pm_{i,j}\right] \approx \arg\left[\mathcal{F}^{-1}\{HP2\, I(u,v)\}\right]; \quad HP2 = |HP_{circ}|HP_{3x3}; \tag{40}$$
$$(e); \quad \left[(-1)^{i+j}\varphi_{i,j} - pm_{i,j}\right] = \arg\left[\mathcal{F}^{-1}\{HP_{circ}I(u,v)\}\right]; \text{ Error-free phase-demodulator.}$$

Being $HP_{circ}(u,v) = e^{i(\pi/4)}[LP_{circ}(u-\pi,v) + LP_{circ}(u+\pi,v)] + e^{-i(\pi/4)}[LP_{circ}(u,v-\pi) + LP_{circ}(u,v+\pi)]$.

Had 4DTC analyzed the frequency content of pixelated interferograms and understood the FTF associated with their SC-PSAs, they could have transformed its 3x3 smooth-cutoff SC-PSA filter into a step-edge quadrature-filter with progressively lower phase error—culminating in an error-free, step-edge quadrature filter solution equivalent to [17].

However, instead of 4DTC building on top of their [9-15], they suddenly turned into Servin's methodology [17],

$$(f); \begin{cases} \dfrac{b}{2}e^{i\varphi} = \mathcal{F}^{-1}\left\{ LP_{circ}\mathcal{F}\left[ I\, e^{-i\,pm} \right] \right\}; & \text{CIO's fringe-carrier product demodulator [17];} \\ \tilde{I}_{alg} = \mathcal{F}^{-1}\left\{ LP_{3x3}\mathcal{F}\left[ I_{det}\, e^{-i\phi_c} \right] \right\}; & \text{4DTC's fringe-carrier product demodulator [18];} \end{cases} \tag{41}$$

To downplay this striking, near-verbatim resemblance to Servin's method, 4DTC deliberately avoided implementing the step-edge Fourier filtering that is necessary to the error-free demodulation strategy [17]. This omission conveniently aligned with the abstract submitted by April 10, 2010 [18], which made no reference to either the fringe-carrier product or low-pass Fourier-domain filtering demodulation. Designing a step-edge filter is trivial and within 4DTC's capabilities—had they intended to use it. The paradox is that this simple yet pivotal component was likely excluded not for technical reasons, but to obscure the methodological borrowing while maintained the functional core of Servin's technique and introducing a superficial variation to deflect potential scrutiny.

4DTC could have also used their own proposed carrier $[\cos(\pi x), \cos(\pi y)]$ [18] for phase demodulation as,

$$(g); \quad e^{i\frac{\pi}{4}}\tilde{I}_{alg} = \mathcal{F}^{-1}\left\{ LP_{3x3}\mathcal{F}\left[ I_{KM}\cos(\pi x) + i\, I_{KM}\cos(\pi y) \right] \right\}; \quad \text{4DTC's carriers as demodulators.} \tag{42}$$

However, it is probable that 4DTC did not know how to apply these carriers to phase-demodulate their proposed simplified pixelated fringe model $I_{KM}(x,y)$; otherwise, they would have taken advantage of their original fringe model variation to deflect the strong methodological overlap with Servin's approach [17].

So, by 2010, there were at least three new distinct phase-demodulation strategies for pixelated fringe demodulation. Given these new and clearly differentiated alternatives—Eq. (40), Eq. (41), and Eq. (42)—it is highly unlikely that both 4DTC and CIO independently converged by 2010 on the same method: the cophased-product Fourier-demodulation approach, represented by Eq. (41)—alternative (*f*). This coincidence is especially improbable considering 4DTC's six-year methodological stagnation with the approaches given in Eq. (39)—alternatives (*a*) and (*b*) [9–15]. To have a graphical picture, we next present three alternatives for pixelated fringe phase demodulation.

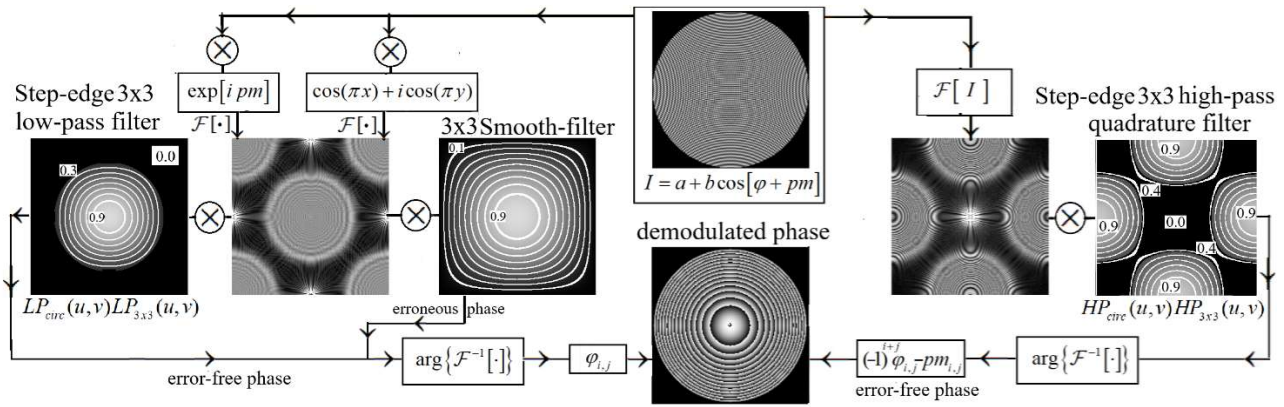

Fig. 24. A schematic view of three alternatives paths for phase-demodulation of pixelated interferograms.

We next present a schematic timeline illustrating the intertwined evolution of 4DTC and CIO's approaches to pixelated phase demodulation, spanning from 2004 to the error-free quadrature Fourier-filtering solution achieved in 2025, as presented in this work. We also show the evolution for the cophased fringe-carrier product phase-demodulation.

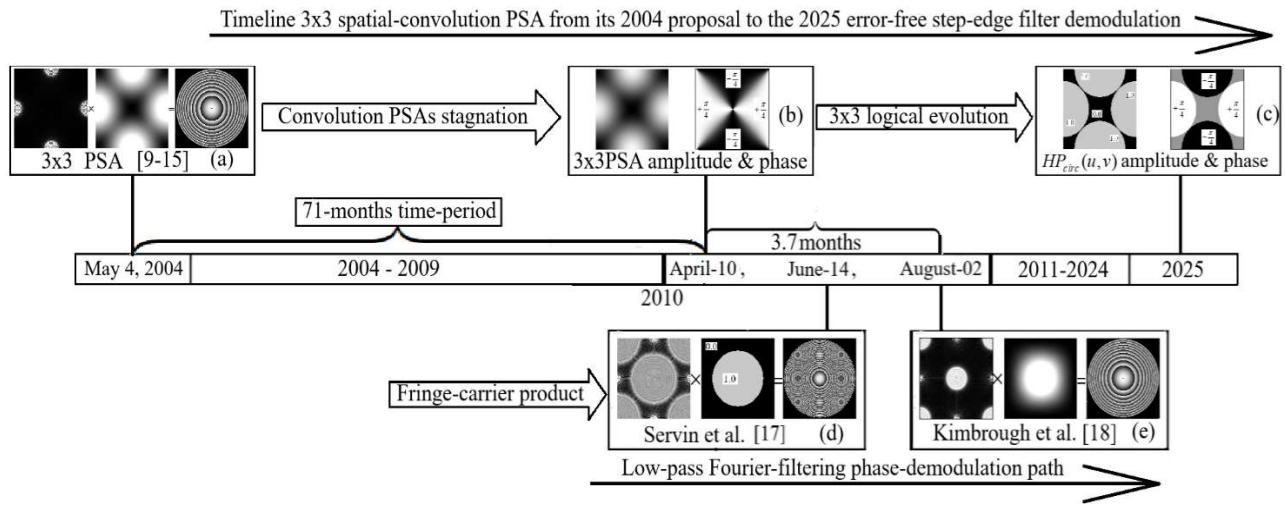

Fig. 25. Top row: Timeline showing the development of pixelated phase demodulation based on the 3×3 SC-PSA from its 2004 patent, US 7,230,717 B2, to its error-free refinement presented in this work. This timeline culminates in 2025 with the correction of the original 4DTC framework [9–15], where the 3×3 smooth-edge high-pass quadrature filter is redefined as a step-edge high-pass quadrature Fourier filter. This correction resolves the long-standing spectral crosstalk issues, completing a methodological path that began in the 2004–2009 period [9–15].
Bottom row: Timeline starting with Servin's 2010 method submitted to Optics Express [17], introducing the cophased fringe-carrier product followed by step-edge low-pass Fourier filtering—submitted just seven weeks before 4DTC's SPIE-7790 presentation [18], which also adopted the cophased product but used a 3×3 kernel smooth-edge low-pass Fourier filter..

Taken together—the six-year lack of spectral analysis, the sudden appearance of Servin's core ideas in [18] just seven weeks after [17] was submitted, and the strategic omission of step-edge filtering—these factors cast serious doubt on the independence of [18]. While parallel development cannot be ruled out entirely, the timing, sequence, and convergence of methods strongly indicate that 4DTC's work in [18] was likely influenced by prior access to Servin's manuscript [17].

## 23. Probability that 4DTC and CIO reached the same fringe-carrier-product Fourier-demodulation

As mentioned, by 2010, there were at least three new and clearly distinct phase-demodulation strategies for pixelated fringe demodulation—Eq. (40), Eq. (41), and Eq. (42). We now estimate the probability that both 4DTC and CIO teams independently arrived at the same cophased fringe-carrier product followed by Fourier low-pass filtering—within just 3.7

months—after 71 months of stagnation by 4DTC using only 2×2 and 3×3 SC-PSAs, as given in Eq. (39)—alternatives (*a*) and (*b*) [9–15].

*25.1 Conjecture 1: Noted discoveries arrive following a Poisson probability distribution,*

$$P(X = k) = \frac{(\lambda t)^k e^{-\lambda t}}{k!} \tag{43}$$

- $P(X = k)$. Probability of *k* discoveries occurring in a time interval (*t*)
- $\lambda$ = Average number of discoveries per month.
- *t* = Length of the time interval in which discoveries may occur.
- *k* = Integer number of discoveries we are calculating the probability for.

From the timeline, we estimate the Poisson's expected value $\lambda$ as,

$$\lambda = \frac{2 \text{ discoveries}}{(71 + 3.7) \text{months}} = 0.0268 \text{ discoveries per month.} \tag{44}$$

The probability of one discovery *k*=1, occurring within 3.7 months after a stagnation of 71 months is:

$$P(X = 1) = \frac{(0.0268 \times 3.7)^{(1)} e^{-(0.0268 \times 3.7)}}{1!} = 0.0898. \tag{45}$$

And the joint probability that 4DTC and CIO make this discovery independently is,

$$P(X = 1) \times P(X = 1) = (0.0898)^2 = 0.0080. \tag{46}$$

The final joint probability that both arrived to the same demodulator having at least 3 possibilities is:

$$P_{indep}[\text{CIO} \cap \text{4DTC}] = [P(X=1)]^2 \times \left(\frac{1}{3}\right)^2 = \left[\frac{0.0898}{3}\right]^2 = 0.00089. \tag{47}$$

Assuming a Poisson distribution to model significant discoveries, one shows that:
- After 71-months of stagnation, the probability of one discovery within 3.7 months is 8.98%
- And if 4DTC and CIO make this discovery independently, lower the chances to 0.8%
- And if both randomly choose the same cophased demodulator among 3 options still lower the chances to 0.089%
- Alternatively, the probability that 4DTC knew Servin's work [17] is 100-0.089=99.91%

*25.2 Conjecture 2: Noted discoveries arrive following a uniform probability distribution,*

From the timeline, the probability of one discovery within $\Delta t = 3.7$ months after a stagnation of $T = 71$ months is:

$$P(\text{one discovery in } \Delta t) = \frac{\Delta t}{T + \Delta t} = \frac{3.7}{71 + 3.7} = 0.0495. \tag{48}$$

And the joint probability that 4DTC and CIO make this discovery independently is,

$$P(X = 1) \times P(X = 1) = (0.0495)^2 = 0.00245. \tag{49}$$

And the final joint probability that both arrive at the same solution having at least 3 possibilities is,

$$P_{indep}[\text{CIO} \cap \text{4DTC}] = [P(X=1)]^2 \times \left(\frac{1}{3}\right)^2 = \left[\frac{0.0495}{3}\right]^2 = 0.000272. \tag{50}$$

Using the uniform probability distribution to model significant discoveries one shows that:
- After 71-months of stagnation, the probability of one discovery within 3.7 months is 4.95%.
- And if both 4DTC and CIO make this discovery independently, lowers the probability to 0.245%.
- And if both randomly choose the same cophased demodulator among 3 options still lower the chances to 0.027%
- Alternatively, the probability that 4DTC knew Servin's work [17] is 100-0.0272=99.97%

This mathematical analysis demonstrates that 4DTC's paper [18], emerging just weeks after Servin's submission [17], as an independent discovery is statistically highly unlikely, and was almost certainly influenced by prior knowledge of Servin's work [17], still under peer review.

**24. Critical need to maximize the demodulation bandwidth in digital pixelated interferometry**

A look at 4D Technology Corporation's (4DTC) web-site reveals compelling reasons for finding an error-free, demodulators for wideband interferograms (https://4dtechnology.com/products/fizeau-interferometers/):
- **High-resolution imaging for steep slopes in aspheres**. The optional 2400 × 2400 high-resolution camera captures the steepest slopes of any commercial interferometer. With its high slope range capability, it enables optical metrology for aspheric optics, freeform surfaces, and highly aberrated elements.

High-slope wavefronts generate wideband interferograms, a critical factor in precision optics, including the fabrication of large astronomical telescopes like the James Webb Space Telescope [21]. For example, a 2400 × 2400 pixelated sensor could theoretically reach a bandwidth of 1697 × 1697 radians/pixels (2400 × 0.7071). This is about the resolution attained by a 1700x1700 pixels, non-pixelated camera mounted in a standard phase-shifting interferometer.

Figure 25 shows a schematic of the Laser Interferometer-Transfer-Function (ITF) chain [18,24]. The coherent imaging system is followed by the pixelated sensor, and finally by the error-free demodulation Fourier algorithm [17].

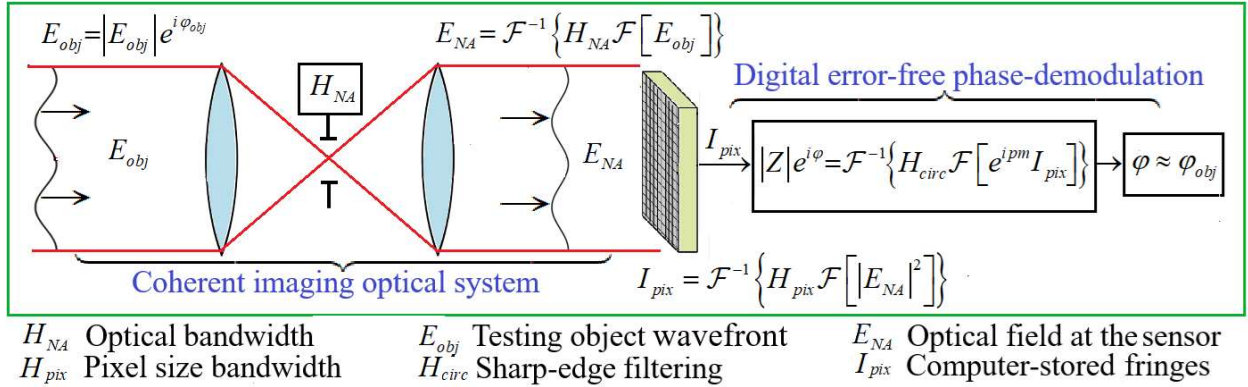

Fig. 26. Opto-digital processing chain. Coherent optical imaging; pixelated sensor; error-free demodulation [18,24].

The previous figure shows the optical filter $H_{NA}$, the pixel-size filter $H_{pix}$, the step-edge filter $H_{circ}$. The bandwidth hierarchy is: $H_{circ} < H_{pix} \ll H_{NA}$. The bottleneck to get the highest testing wavefront-bandwidth comes from $H_{circ}$.

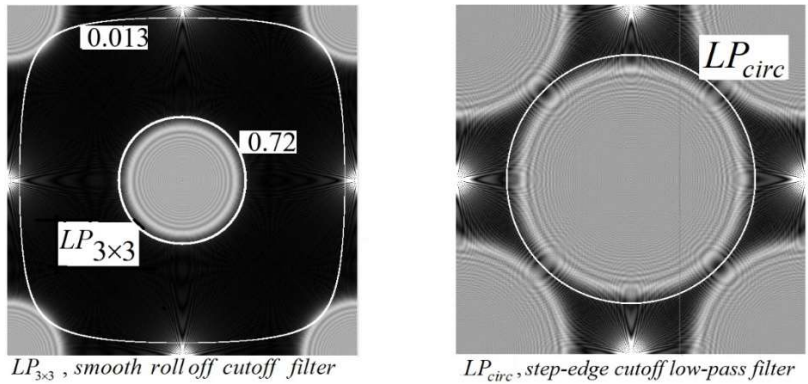

Fig. 27. Fringe-carrier product Fourier-interferometry [17,18]. Servin's bandwidth-gain using a step-edge filtering [17] with respect to the 3x3 smooth-edge low-pass filter for about the same small phase error [18].

A recent historical account of optical metrology's role in constructing the James Webb Space Telescope was presented by Miller et al. —4D Technology Corporation— in their article, "The Metrology Story Behind the James Webb Space Telescope," published in Laser Focus World (July 10, 2022) [21]. Surprisingly, the article omits any mention of the development of error-free digital phase demodulation of pixelated interferograms [17]—a crucial advancement that directly impacts the performance of 4DTC's groundbreaking Fizeau pixelated interferometer [9]. This omission is particularly notable given that digital demodulation is the bandwidth bottle-neck of the opto-electronics-digital processing chain in a pixelated interferometer. This performance reached their maximum digital bandwidth since 2010, following Servin's discovery of error-free Fourier cophased phase demodulation [17].

The omission of Servin et al.'s seminal paper, "*Error-free demodulation of pixelated carrier frequency interferograms*" [17], by 4DTC is not an isolated lapse—as far as we know, it has been systematic across all of their post-2010 publications on pixelated interferometry. Despite the clear and substantial impact of Servin's methodology—which enabled the full measurement capabilities of 4DTC's principal line of commercial pixelated interferometers—as far as we know, this foundational contribution has been consistently and deliberately kept uncited in all subsequent 4DTC literature. This pattern appears calculated to obscure Servin's prior work [17], a fact easily corroborated through a simple Google Scholar search. The facts are unequivocal: after six years of methodological stagnation with their flawed 2×2 and 3×3 spatial convolution PSAs, the definitive solution emerged externally. Although acknowledging this breakthrough may have been professionally uncomfortable for 4DTC, ethical standards unequivocally mandate proper attribution of foundational work. Their sustained failure to credit this advance undermines scholarly transparency, falsifies the historical account of innovation, and represents a clear violation of academic integrity.

## 25. Discussion

We summarize circumstantial evidence which suggest that 4DTC [18] may have relied on Servin's key ideas [17]:

- Between 2004 and 2009, 4DTC was the only institution working on the demodulation of pixelated interferograms, as their own references in [18] show. In 2010, Servin's group at the CIO in Mexico entered the field [16,17], and on June 14, submitted to Optics Express the definitive, error-free solution to this six-year unresolved problem [17]. Strikingly, just seven weeks later, on August 2, 2010, K&M (4DTC) presented their paper [18] at the SPIE-7790 conference—a presentation exhibiting unmistakable methodological parallels with Servin's approach [17]. At that time, 4DTC and Servin's team were the only groups worldwide pursuing an error-free demodulator for pixelated interferograms [17,18]. Given the timing, the overlap in objectives, and the evident similarities in methodology, it is highly likely that Kimbrough and Miller had access to Servin's manuscript during the peer-review process and incorporated key aspects of it into their own work [18].
- Surprisingly, before 2010, none, including 4DTC [9-15], had published a mathematical model for pixelated interferograms, nor its Fourier spectrum, neither the fringe-carrier product spectrum that immediately unlocks the correct demodulation of pixelated fringes [17].
- The abrupt shift from their previously flawed high-pass 2×2 and 3×3 spatial-convolution PSAs (SC-PSAs) [9–15] to their "new approach" in [18]—merely a repackaging their high-pass 2x2 and 3x3 kernels to 2x2 and 3x3 low-pass kernels—was left completely unaddressed and unexplained in their 2010 paper [18].
- In their 2010 SPIE-7790 presentation, Kimbrough and Miller state: *"A pixelated mask is another method of producing a carrier phase. Fundamentally the only difference between pixelated mask spatial carrier and other spatial carrier methods is the form of the carrier wavefront"* (italics mine) [18]. If the difference between pixelated and linear carriers is so fundamentally minor, one must ask: why did 4DTC fail to acknowledge this equivalence throughout their publications from 2004 to 2009 [9–15]? Why was this realization only articulated in 2010 [18], just weeks after Servin et al. submitted their manuscript to *Optics Express* [17]? The timing and sudden conceptual shift raise valid concerns about whether this recognition was independently derived or prompted by prior exposure to Servin's still in peer-review manuscript.
- In K&M's paper [18], there is a striking disconnect between the title, abstract, and keywords submitted before April 10, 2010, and the body of the paper ultimately presented at SPIE-7790 on August 2, 2010. The hearth concept of the paper's body [18] is: The fringe-carrier complex product, and low-pass Fourier-demodulation [18]. Surprisingly, neither of these concepts, appears in the title, abstract, or keywords in [18]. Instead, the title, abstract and keywords continue to focus on 2x2 and 3x3 spatial-processing convolution PSAs [18].
- Furthermore, 4DTC presented another paper at the same SPIE-7790 conference [19]. James Miller coauthored both papers [18,19]. Unexpectedly, in [19], there is no mention of K&M's new fringes-carrier product, Fourier demodulation presented in the very same SPIE conference [18]. Instead, paper [19] repeats a paragraph in US-7,230,717B2 [9]: "*An preferred method for calculating the phase difference at each spatial coordinate is to combine the measured signals of neighboring pixels in a fashion similar to a windowed convolution algorithm. This method provides an output phase-difference map having a total number of pixels equal to (n-W) times (m-V), where W and V are the sizes of the correlation window in the X and Y directions, respectively*." The grammatical error "An preferred" is in [19]. Thus, by April 10, 4DTC was still relying on SC-PSAs.
- Kimbrough and Miller (K&M)) stated: "*If a symmetric filter with a step edge cutoff were possible, the dashed line in each spectrum indicates the maximum spatial frequency to prevent aliasing for a wavefront with only tilt and a flat background intensity*" (italics mine) [18]. The conditional phrasing—"*if ... were possible*"—erroneously suggests that implementing a step-edge Fourier filter is difficult or unfeasible. In reality, any practitioner with basic knowledge in digital interferometry recognizes that coding a step-edge Fourier filter is both elementary and routine. This inexistant difficulty strongly suggests that Kimbrough and Miller deliberately sidestepped incorporating such a filter—despite its trivial implementation—in a calculated effort to portray their method in [18] as methodologically distinct from Servin's explicit, step-edge-based solution [17].
- Had K&M's 2010 paper [18] been a truly original development, it is reasonable to expect that 4DTC would have fully disclosed their methodological shift transparently and explicitly. For example, they might have written:
    > *"The smooth roll-off cutoff frequency response of the 2×2 and 3×3 spatial convolution phase-shifting algorithms (SC-PSAs) utilized between 2004 and 2009 consistently produced phase detuning errors [9-15]. It is now recognized that these errors stem from spectral crosstalk between the complex conjugate components of the filtered pixelated interferogram—components that ideally must remain separated. The*

*gradual spectral attenuation intrinsic to 2x2 and 3x3 smooth-cutoff filters failed to prevent the mixing of these conjugate spectra, thereby introducing systematic phase-demodulation errors. This issue was particularly pronounced when applied to wideband interferograms, where this wide bandwidth of the fringes exacerbated the overlap between these ideally-separated, conjugate spectral components.*

*To address this critical issue, we now introduce the fringe-carrier synchronous-cophased product demodulation strategy that shifts the desired analytic signal to the spectral origin while simultaneously relocating its conjugate component to the high-frequency region. This newly presented fringe-carrier product allows the use of a step-edge low-pass Fourier-filter, which replaces the previously employed 2x2 and 3x3 smooth-edge high-pass filters [9–15]. This sequence of fringe-carrier product followed by step-edge low-pass filtering ensures to fully delete the conjugate signal, thereby eliminating phase detuning errors caused by previous smooth-edge filters and finally enabling truly error-free phase-demodulation—even under wideband fringe conditions. This new methodology represents a major conceptual breakthrough, effectively resolving the six-year limitations that had constrained the bandwidth, and overall phase-demodulation performance of commercialized 4DTC's pixelated interferometers [9–15]."*

Such a transparent, clear and logical explanation is notably absent from [18], reinforcing the concern that the conceptual and paradigm-shift may not have originated independently within 4DTC, but rather as a consequence of prior exposure to Servin's manuscript still in peer-review [17].

- Had 4DTC analyzed the frequency spectra of pixelated interferograms and correctly interpreted the FTF of their smooth-cutoff 2x2 and 3×3 spatial convolution PSAs (SC-PSAs), they could have constructed a sequence of step-edge quadrature filters with progressively lower phase-error. This process—just requiring thresholding to the frequency response of their existing SC-PSAs—would have enabled them to construct a sequence of improved sharper-edge quadrature Fourier filters, ultimately leading to an error-free step-edge quadrature filter fully equivalent to the Fourier-based demodulation approach introduced by Servin et al. [17].

    Crucially, such a solution could have emerged entirely from within 4DTC's previously established mathematical framework [9–15], requiring no conceptual departure from their long-standing methodology. It would not have necessitated the sudden adoption of an external paradigm—namely, the fringe-carrier product and Fourier low-pass filtering proposed in Servin's manuscript [17]. The fact that 4DTC overlooked this direct and technically accessible evolutionary path, and instead converged so closely with Servin's formulation just weeks after his submission, strengthens the suspicion that 4DTC's 2010 paper [18] may have drawn heavily on Servin's ideas. In particular, the similarity suggests that Servin's work influenced [18], even though this paper cites no such source.

Finally, it is essential to acknowledge that 4D Technology Corporation (4DTC) is a well-established U.S.-based manufacturer of laser optical interferometers, which are commercially distributed worldwide for applications in optical metrology. In addition, 4DTC supplies the proprietary software used for phase demodulation of the interferograms produced by their instruments https://4dtechnology.com/products/fizeau-interferometers/ .

- A plausible motivation for 4D Technology Corporation to have drawn upon the still-unpublished manuscript by Servin et al. [17] may have been the strategic interest in presenting their work as an independent development of a novel phase-demodulation method with potential commercial value—thereby avoiding potential patent-related conflicts with the CIO (despite the fact that the CIO had no intention of pursuing a patent). If 4DTC had access to Servin's manuscript during the peer-review process, they would have had a some weeks window to incorporate key innovations from [17], particularly the cophased fringe-carrier product and Fourier-domain low-pass filtering, while deliberately omitting explicit reference to the step-edge filter. This approach enabled 4DTC to publicly position their work as a parallel advancement in Fourier-based pixelated interferometry, despite having previously dismissed Fourier demodulation in their publications from 2004 to 2009 [9–15]. In summary, 4DTC's 2010 method closely parallels the technique introduced by Servin et al. [17]. The convergence is substantial, yet [18] offers neither a rationale for this methodological shift nor acknowledgment of the prior work [17].

## 26. Conclusion

In this work, we have reviewed the evolution of phase demodulation techniques for pixelated interferograms during the period 2004–2010. Our analysis identifies serious methodological limitations in the 2×2 and 3×3 spatial-processing phase-shifting algorithms (SC-PSAs) developed and employed by 4DTC between 2004 and 2009 [9–15]. These algorithms, based on 4-step and 9-step temporal PSA formulas, spatially arranged into complex convolution kernels, were systematically applied without any Fourier-domain spectral analysis throughout that period [9–15].

   We begin this paper by noting that the foundational Ansatz in Servin et al.'s manuscript [17] was the hypothesis that pixelated-carrier interferometry possesses a structural mathematical analogy to linear-carrier interferometry. This conjecture was confirmed obtaining error-free phase demodulation [17]. In stark contrast, without prior indication, rationale, or continuity with their 2004–2009 research [9–15], K&M abruptly asserted in their 2010 SPIE-7790 paper [18]: "*A pixelated mask is another method of producing a carrier phase. Fundamentally, the only difference between pixelated mask spatial carrier and other spatial carrier methods is the form of the carrier wavefront*" (italics mine).

Why did 4DTC choose to acknowledge and articulate this ostensibly elementary fact only in 2010 [18], after systematically omitting any reference to it over six years of prior publications [9–15]? The most plausible explanation is that, by that time, 4DTC had accessed Servin's manuscript [17] as the most technically qualified peer-reviewing institution, positioning them to appropriate this foundational insight. The language in [18] appears carefully engineered to retroactively normalize the concept, thereby diminishing both the originality and the significance of Servin's contribution while fostering the false impression that this equivalence had long been established knowledge. This abrupt recognition, following six years of omission, stands in direct contradiction to 4DTC's persistent failure to identify or disclose this fundamental equivalence between 2004 and 2009 [9–15].

Continuing our exposition, we apply Fourier analysis to rigorously demonstrate that the 2×2 and 3×3 SC-PSAs function as spatial high-pass quadrature filters—an essential technical fact that 4DTC systematically failed to acknowledge, investigate, or disclose in any of their pre-2010 publications [9–15].

More concerning is that the phase errors inherent in 4DTC's 2×2 and 3×3 SC-PSAs, used from 2004 to 2009 [9–15], persisted unchanged into their 2010 SPIE-7790 presentation [18]. Rectifying their persistent phase demodulation error—spanning now a full seven-year period (2004–2010)—would have required a straightforward substitution of their spatial convolution smooth-edge kernel-based filters by a step-edge cutoff low-pass Fourier filter. Instead, in [18], K&M wrote: "*If a symmetric filter with a step edge cutoff were possible, the dashed line in each spectrum indicates the maximum spatial frequency to prevent aliasing for a wavefront with only tilt and a flat background intensity*" (italics mine). This statement misleadingly suggests that constructing a step-edge Fourier filter is impractical or unfeasible. Technically, this is incorrect; implementing such a filter is trivial within standard digital signal processing frameworks.

The timeline and circumstantial evidence presented here point to a deliberate and strategic decision by 4DTC to avoid implementing step-edge low-pass Fourier filtering, in order to maintain alignment with the abstract submitted for SPIE-7790 prior to April 10, 2010 [18]. That abstract still promoted their legacy, technically flawed 2×2 and 3×3 spatial-convolution phase-shifting algorithms (SC-PSAs) [9–15]. Including the step-edge Fourier filter would have further highlighted their unacknowledged pivot to the fringe-carrier product Fourier demodulation approach introduced by Servin et al. [17]. By that point, it is highly plausible that 4DTC had already reviewed Servin's manuscript—which presented the definitive, error-free method based on this exact filtering strategy. Their omission of this simple yet crucial technique appears calculated to obscure their reliance on Servin's intellectual contribution. Notably also, 4DTC failed to disclose that their newly rebranded 2×2 and 3×3 low-pass kernel filters [18] exhibit the very same demodulation errors as their earlier previous high-pass counterparts [9-15].

These observations raise important questions about 4DTC's portrayal of their work [18] as an independent discovery, suggesting a need for closer examination of the conceptual and mathematical foundations underlying their claims. This documented exposé will remain publicly accessible, available to anyone willing to trace the appropriation of core mathematical concepts from Servin et al. manuscript [17]; submitted just seven weeks before 4DTC's presentation [18]. The timing and presentation choices imply that 4DTC may have sought to reduce exposure to potential intellectual property concerns with CIO, effectively positioning their approach to avoid explicit acknowledgment of Servin's prior contributions [17].

We have herein demonstrated that the pivotal breakthrough in this historical sequence was the genuinely independent contribution by Servin et al.: the introduction of the pixelated fringe-model, the synchronous-cophased fringe-carrier product and the step-edge low-pass Fourier filtering [17]—achieving, for the first time, error-free phase demodulation of wideband pixelated-carrier interferograms. This innovation conclusively resolved the seven-year-long open problem in digital pixelated interferometry, which had remained unsolved from 2004 through 2010 [9–15,18].

Notably, a 2022 historical review by 4DTC, "The Metrology Story Behind the James Webb Space Telescope" (Laser Focus World [21]), omits any reference to Servin et al.'s pivotal development in pixelated digital interferometry [17]. This omission is significant given that error-free phase demodulation of pixelated fringes is fundamental to the performance of 4DTC's flagship Fizeau pixelated interferometer [9], a core technology employed in the optical metrology of the James Webb Space Telescope's optics [21]. Since 2010, 4DTC has systematically excluded citation of Servin et al.'s seminal paper [17] across all subsequent publications on pixelated interferometry, despite its decisive role in enabling unlocking the full digital bandwidth of their commercialized interferometers. This sustained omission constitutes a clear violation of academic integrity and distorts the documented record of technological innovation.